\pdfoutput=1

\documentclass[aps,prd,twocolumn,superscriptaddress]{revtex4-1}
\usepackage{graphicx}
\usepackage{booktabs}
\usepackage[hidelinks]{hyperref}
\usepackage{color}
\hypersetup{
    colorlinks,
    linkcolor={black},
    citecolor={black},
    urlcolor={black}
}
\usepackage{amssymb,amsmath,amsfonts}
\begin{document}


\title{Scaling violation and the magnetic equation of state in chiral models}

\author{G\'abor Andr\'as Alm\'asi}
\email[]{g.almasi@gsi.de}
\affiliation{Gesellschaft f\"{u}r Schwerionenforschung, GSI, D-64291 Darmstadt, Germany}
\affiliation{Technische Universit\"{a}t Darmstadt, D-64289 Darmstadt, Germany}
\author{Wojciech Tarnowski}
\affiliation{Faculty of Physics, Astronomy and Applied Informatics, Jagiellonian University, PL-30-348 Cracow, Poland}
\author{Bengt Friman}
\affiliation{Gesellschaft f\"{u}r Schwerionenforschung, GSI, D-64291 Darmstadt, Germany}
\author{Krzysztof  Redlich}
\affiliation{ExtreMe Matter Institute EMMI, D-64291 Darmstadt, Germany}
\affiliation{Faculty of Physics and Astronomy, University of Wroc\l aw, PL-50-204 Wroc\l aw, Poland}
\affiliation{Department of Physics, Duke University, Durham, North Carolina 27708, USA}

\begin{abstract}
The scaling behavior of the  order parameter at the chiral phase transition, the so-called magnetic equation of state, of strongly interacting matter is studied within effective  models.
We  explore universal and  nonuniversal structures  near the critical point. These include the  scaling functions, the leading corrections to scaling  and the corresponding size of  the scaling window as well as their dependence on an external symmetry breaking field.   We consider two models in the mean-field approximation, the  quark-meson and the Polyakov loop extended quark-meson (PQM) models,  and compare  their critical properties with  a purely bosonic theory, the $O(N)$ linear sigma model in the $N\to\infty$ limit. In these models the order parameter scaling function is found  analytically using  the  high temperature expansion of the  thermodynamic potential.  The effects of a gluonic  background on the nonuniversal scaling parameters are studied within the PQM model.
\end{abstract}


\maketitle

\hypersetup{
    colorlinks,
    linkcolor={blue},
    citecolor={blue},
    urlcolor={blue}
}
\section{Introduction \label{sec:Introduction}}

Spontaneous  chiral symmetry breaking and its restoration at finite temperature and density is an essential ingredient in our understanding  of the phase structure of strongly interacting matter and hence a key problem in QCD~\cite{Friman:2011zz,Fukushima:2010bq,Fukushima:2013rx}.

In the limit of  massless light quark flavors, the chiral phase transition in quantum chromodynamics (QCD)  was conjectured to be of second order, in  the
$O(4)$ universality class~\cite{Pisarski:1983ms}. Current lattice QCD (LQCD)
simulations at physical  up, down and strange quark masses show that at vanishing and small baryon density the
transition from a hadron gas to a quark gluon plasma is a smooth crossover \cite{Aoki:2006we}. Moreover, lattice studies of the scaling properties of the chiral order parameter are consistent  with the conjectured $O(4)$ symmetry and indicate that the scaling violations are fairly small for physical quark masses~\cite{Ejiri:2009ac,Kaczmarek:2011zz,Ding:2013lfa,Ding:2015pmg}. Consequently, quantities that are sensitive to chiral criticality, are expected to exhibit characteristic
properties governed by the universal singular part of the free energy density. The magnetic
equation of state, which reveals the scaling of the chiral order parameter as a function
of the reduced temperature and the quark masses,  is a key  quantity in this context \cite{zinn2002quantum}. We note, however, that the issue whether the chiral transition of QCD exhibits $O(N)$ scaling is quite subtle. Indeed, several studies suggest that in the chiral limit the transition could be first order \cite{D'Elia:2005bv,Bonati:2009zz,Bonati:2014kpa}.

The critical properties   of QCD are  often studied in effective models that share the   chiral symmetry of the QCD Lagrangian and exhibit spontaneous breaking of this symmetry in vacuum. Popular models include the quark-meson  (QM) model~\cite{GellMann:1960np} and its Polyakov loop extended version (PQM)~\cite{Schaefer:2007pw,Schaefer:2009ui,Mizher:2010zb,Herbst:2010rf,Skokov:2010wb,Skokov:2010sf,Skokov:2010uh,Mitter:2013fxa,Herbst:2013ufa},  the  $O(N)$ linear sigma (LS) model~\cite{AmelinoCamelia:1992nc,AmelinoCamelia:1997dd,Petropoulos:1998gt,Lenaghan:1999si,Lenaghan:2000ey,Jakovac:2003ar,Andersen:2004ae,Andersen:2008qk} as well as the Nambu-Jona--Lasinio model~\cite{Nambu:1961tp,Nambu:1961fr,Fukushima:2003fw,Ratti:2005jh,Sasaki:2006ww,Roessner:2006xn,Fukushima:2008wg}. In  the chiral limit, all these models undergo a second-order phase transition of the $O(4)$ universality class.  Consequently, they belong to the same  universality class as QCD.

The nonzero $u$ and $d$ quark masses  break the chiral symmetry explicitly. However,  for  small masses, the dynamics is by and large determined by the underlying second-order phase transition, while the nonzero quark masses act  as a weak perturbation. Clearly, even at small  masses,  there is  no phase transition in a strict sense.  In a crossover region, the order parameter decreases smoothly from a large value at small temperatures and densities to a very small but finite one at high temperatures and densities. The melting of the order parameter near the critical point is captured by the magnetic equation of state.

The value  of the light  quark mass up to which the critical fluctuations of the  underlying second-order phase transition dominate the physics near the pseudocritical point is model dependent and consequently nonuniversal. As noted above, LQCD calculations suggest ~\cite{Ejiri:2009ac,Kaczmarek:2011zz,Ding:2013lfa,Ding:2015pmg}, that for physical  quark masses the critical behavior of the chiral condensate is well approximated by the $O(N)$ scaling magnetic equation of state. This indicates that the scaling window of the QCD chiral crossover transition extends more or less to the physical light quark masses.

In functional renormalization group (FRG)~\cite{Wetterich:1992yh,Morris:1993qb,Berges:2000ew,Polonyi:2001se} studies of the QM model, it was shown \cite{Braun:2010vd}   that  at the  physical pion mass,   the behavior of the condensate is not well described  by the universal scaling function, in spite of the fact that in the chiral limit  this theory belongs to the $O(4)$ universality class. The different critical behavior of  QCD and the QM model within the FRG approach is linked to the scaling breaking terms in the magnetic equation of state,  which are nonuniversal.

In  this paper,  we  explore  the critical properties of the chiral order parameter and the magnetic equation of state in QCD-like models. We  focus on their universal and  nonuniversal structure  near the critical point. This includes a derivation  of the  scaling functions and  leading-order scaling violating corrections. In particular, we  systematically study the dependence of the magnetic equation of state on an external symmetry breaking field, and assess the size of the critical region. For transparency, we consider the QM  and PQM models in the mean-field approximation, where only fermionic fluctuations are accounted for, and confront  their critical properties with  a purely bosonic theory, the $O(N)$ linear sigma model. In the mean-field approximation to the QM model as well as in the $N\to\infty$ limit of the LS model, the calculation of the magnetic equation of state is  carried out analytically by employing the high temperature expansion.
The effects of the gluonic  background on the nonuniversal scaling parameters are assessed in the PQM model.

We stress that although these models do not reproduce the expected scaling behavior of QCD on a quantitative level, they provide a transparent framework for exploring chiral criticality. Moreover, this study yields new insight into possible patterns of scaling violation exhibited by the magnetic equation of state.

The paper is organized  as follows. In Sec. \ref{sec:Universality} we briefly summarize the theory of second-order phase transitions and  introduce the magnetic equation of state. In Sec. \ref{sec:Landau-theory} the magnetic equation of state is discussed within Landau theory. The effective Landau coefficients are obtained in Sec. \ref{sec:QM} for the QM model. The LS model and its magnetic equation of state are introduced in Sec. \ref{sec:Oinfintro}. In Sec. \ref{sec:mEoS} we compare  results for  magnetic equation of state in  different models and discuss the nonuniversal corrections. In the final section,  we present a summary and conclusions.

\section{Universality and scaling \label{sec:Universality}}
In the scaling theory of phase transitions,  the
free energy density $f(T,H)$ is, in the vicinity of a second-order critical point,
split into a singular scaling part $f_s(T,H)$ and a regular part. In a given  universality class, the singular part has a  universal structure \cite{zinn2002quantum}.

For a given temperature $T$ and external field $H$, one introduces the  scaling variables
\begin{equation}\label{eq:scal-var}
	\overline{t}=\frac{t}{t_0}=\frac{T-T_c}{T_c t_0}, \quad h=\frac{H}{H_0 h_0}, \quad z=\frac{\overline{t}}{h^{1/(\beta\delta)}},
\end{equation}
where $T_c$ is the critical temperature and $t_0$, $H_0$ and $h_0$ are appropriately chosen constants. In terms of these variables,  the scaling part of the free energy has   the universal form
\begin{equation}\label{universal}
	f_s(T,H) = F_0\; h^{1+1/\delta} f_f\left(z\right).
\end{equation}
The scaling of the order parameter $\langle\sigma\rangle$ is
obtained from  Eq.  (\ref{universal}),
\begin{equation} \label{eq:ScalingOP}
	\langle\sigma\rangle = \frac{\partial}{\partial H} f_s(T,H) = \sigma_0 h^{1/\delta} f_G(z).
\end{equation}
In Eqs. (\ref{universal}), (\ref{eq:ScalingOP}) $F_0$ and $\sigma_0$ are again appropriately chosen constants.
The functions $f_f$ and $f_G$ and the critical exponents are universal, as they do not depend on the details of the model, but only on its universality class. The scaling function $f_G$ has the following asymptotic properties: $f_G(0) = 1$ and $\lim_{z\rightarrow -\infty} \frac{f_G(z)}{(-z)^\beta} = 1$.

From the scaling  function, one arrives at  the following  well-known  scaling properties of the order parameters on the coexistence line ($
T <T_c, H=0$) and at the pseudocritical point ($T=T_c, H>0 $):
\begin{equation}\label{eq:h0t0def}
\langle{\sigma}\rangle = \left\{\begin{array}{lr} \sigma_0 h^{1/\delta} = \sigma_0 \left(\frac{H}{H_0 h_0}\right)^{1/\delta}, &  T=T_c,\; H>0 \\
\sigma_0 \left(-{\overline{t}}\right)^{\beta}~~~ = \sigma_0 \left(\frac{(-t)}{t_0}\right)^{\beta}, &
H=0,\;
T<T_c \end{array}\right..
\end{equation}
The normalization constants $t_0$ and $h_0$ are determined by these equations, once $\sigma_0$ and $H_0$ are specified. We choose $\sigma_0 = \left\langle \sigma \right \rangle_{T=0} = f_{\pi}\approx 93 \;\mathrm{MeV}$ and $H_0=m_{\pi}^2f_{\pi}\approx 1.77\times 10^6\;\mathrm{MeV}^3$.

The scaling of the order parameter susceptibility in the vicinity of  the critical point is obtained  from Eq.  (\ref{eq:ScalingOP}),
\begin{equation}
\resizebox{.48 \textwidth}{!} {$
	\chi_{\sigma} = \frac{\partial \langle\sigma\rangle}{\partial H}  =
	\chi_0 h^{1/\delta-1}\left( f_G(z) - \frac{z f_G'(z)}{\beta} \right)
	\equiv \chi_0 h^{1/\delta-1} f_\chi(z).
$}
\end{equation}
Consequently, the maximum of the susceptibility is located at a fixed value of $z=z_p$, independently of the external field $H$. From this, it follows  that the  pseudocritical temperature at a  finite external field is  given by~\cite{Ejiri:2009ac}
\begin{equation} \label{eq:Tpcdependence}
	\frac{T_p(H)-T_c}{T_c} = \frac{z_p}{z_0} \left(\frac{H}{H_0}\right)^{1/(\beta\delta)},
\end{equation}
where $z_0 = {h_0^{1/(\beta\delta)}}/{t_0}$  is a nonuniversal parameter.

 The width of the crossover region can be  defined from the susceptibility of the order parameter $\chi_{\sigma}$. The universal part of $\chi_{\sigma}$ is a peaked function with a width of $\Delta z$, which  depends only on the universality class. Thus, the width of the crossover region in temperature, which for a given external field given by
\begin{equation} \label{eq:crossoverwidth}
	\frac{\Delta T}{T_c} = \frac{\Delta z}{z_0} \left(\frac{H}{H_0}\right)^{1/(\beta\delta)},
\end{equation}
depends on the nonuniversal parameter $z_0$.

\section{Modeling magnetic equation of state  \label{sec:modelling} }
The scaling theory of phase transitions provides definite predictions for the critical properties of various thermodynamic observables. These are characterized by the critical exponents of the corresponding universality class, which are ingrained in the scaling free energy. In particular,  the order parameter  is given by the magnetic equation of state  ${\langle\sigma\rangle}/(\sigma_0 h^{1/\delta})=x(z,h)$,  which in the critical region collapses to the  universal  scaling function  $x=f_G(z)$, introduced in Eq. (\ref{eq:ScalingOP}).  However, sufficiently far away from the critical point, corrections to the universal scaling become significant and $x(z,h)$ deviates from $f_G(z)$. The size of the scaling region is not universal, and hence model dependent.

In the following, we focus on the QCD chiral phase transition and discuss the  scaling properties of the chiral order parameter near the  critical point.  We consider effective models  belonging  to the universality class of the QCD chiral transition,  and  compute  leading-order corrections to the scaling curve in the magnetic equation of state. We analyze scaling violations induced by finite quark masses in the context of recent LQCD findings, which indicate that for a physical value of the pion mass, QCD lies in the scaling regime  of the underlying second-order phase transition. We consider  the  QM   and  PQM models in the mean-field approximation   and  a purely bosonic theory, the $O(N)$ linear sigma model,  in the $N\to\infty$ limit. We examine, to what extent the scaling violating terms are compatible with QCD for a physical value of the pion mass.

In the next section   we consider the  Landau theory of  second-order  phase transitions and construct the corresponding magnetic equation of state  as a baseline for a quantitative description of the QM and PQM models. We then go beyond the mean-field approximation and study the magnetic equation of state and deviation from scaling  in the large-$N$ limit of the  $O(N)$ linear sigma model, using the high temperature expansion.

\subsection{The Landau theory \label{sec:Landau-theory} }

In  mean-field theory,  second-order phase transitions are generically described by Landau theory \cite{landau2013statistical}. There, the effective potential is a polynomial in the order parameter $\sigma$, with coefficients that are analytic functions of the temperature $T$. Assuming a symmetry under reflections, $\sigma \rightarrow -\sigma$, the effective potential is, apart from a symmetry breaking term proportional to the external field $H$, an even polynomial in $\sigma$, and reads
\begin{equation}\label{landau}
	\mathcal{L}(T,H;\sigma) =  a(t)\frac{\sigma^2}{2}+
	b(t)\frac{\sigma^4}{4}+
	c(t)\frac{\sigma^6}{6} +
	d(t)\frac{\sigma^8}{8} + \dots -H\sigma.
\end{equation}
Here the $T$-dependent coefficients are parameterized as polynomials in the reduced temperature $t=\frac{T}{T_c}-1$ and have the form
\begin{align}
\begin{split}
	a(t)&=a_1 t + a_2 t^2 + a_3 t^3 + \dots, \\
	b(t)&=b_0 + b_1 t + b_2 t^2 + \dots, \\
	c(t)&=c_0 + c_1 t + \dots, \\
	d(t)&=d_0 + \dots .
	\label{eq:params}
	\end{split}
\end{align}

For a given value of $T$ and $H$,  the order parameter is given by the location of the minimum of $\mathcal{L}(T,H;\sigma)$. This is determined by solving the gap equation
\begin{equation} \label{eq:Landau_gapeq}
\resizebox{.50 \textwidth}{!} {$
	 \frac{\partial \mathcal{L}}{\partial \sigma}=a(t) \left\langle\sigma\right\rangle+
	b(t) \left\langle\sigma\right\rangle^3+
	c(t) \left\langle\sigma\right\rangle^5 +
	d(t) \left\langle\sigma\right\rangle^7 + \dots = H.
$}
\end{equation}
\resizebox{.48 \textwidth}{!} {For vanishing external field and $b(t)>0$ and $c(t),d(t)\geq 0$,} there is a second-order phase transition at $T=T_c$, where $a(t)$ vanishes.
Around the corresponding critical point, the order parameter
exhibits   the following     scaling properties:
\begin{equation}
\langle{\sigma}\rangle = \left\{\begin{array}{lr} \left(\frac{H}{b_0}\right)^{1/\delta}, &  T=T_c,\; H>0 \\
\left({\frac{a_1(-t)}{b_0}}\right)^{\beta}, &
H=0,\;
T<T_c \end{array}\right. ,\label{eq14}
\end{equation}
with $\delta=3$ and $\beta=1/2$, respectively.
Comparing Eq. (\ref{eq14})   with the general scaling behavior of the order parameter in Eq. (\ref{eq:h0t0def}) one can extract  $t_0$ and $h_0$ in Landau theory,
\begin{equation} \label{eq:MF_t0h0}
	h_0 = \frac{b_0 \sigma_0^3}{H_0}, \quad \quad  t_0 = \frac{b_0 \sigma_0^2}{a_1}.
\end{equation}
We note that both $h_0$ and $t_0$ depend on the normalization of the order parameter, while $h_0$ depends also on the normalization of the external field. Thus, a comparison of these parameters between  different models has to be done with care. The model dependence can be reduced by considering the combination $z_0 = h_0^{2/3}/t_0 = \frac{a_1 b_0^{2/3}}{b_0 H_0^{2/3}}$, which is independent of $\sigma_0$. Hence, one can compare the value of $z_0$ with other approaches, provided the normalization of the external field $H_0$ is known.

The mean-field magnetic equation of state is obtained from Landau's thermodynamic potential, by introducing the scaling variables
\begin{align} \label{eq_Landau_xzdef}
\begin{split}
	x&=\frac{\sigma / \sigma_0}{h^{1/3}} = \sigma \left( \frac{b_0}{H} \right)^{1/3}, \\
	z&=\frac{\overline{t}}{h^{2/3}} = \frac{a_1 t}{b_0}\left(\frac{b_0}{H}\right)^{2/3},
\end{split}
\end{align}
where $x>0$ and $z$ can take any real value. The variable $z$ can be used to map out the phase diagram of the system.
Thus, $\left|z\right|\ll 1$ corresponds to a system near the critical point $T=T_c$, while $z\ll -1$ refers to the phase with broken symmetry and $z\gg 1$ to the one where the symmetry is restored.

Using Eq.  (\ref{eq_Landau_xzdef}),  we express the reduced temperature and the order parameter in terms  of $x,z$ and $H$,
\begin{equation}\label{eq17}
	t= \frac{b_0 z}{a_1}\left(\frac{H}{b_0}\right)^{2/3}=\frac{z}{z_0}\left(\frac{H}{H_0}\right)^{2/3} ,
	\quad\quad \sigma = x \left( \frac{H}{b_0} \right)^{1/3}.
\end{equation}
The gap equation  Eq. (\ref{eq:Landau_gapeq}) can be expressed in terms of the scaling variables $x,z$
\begin{align} \label{eq:Landau_mEOS}
	\left(x(x^2+z)-1\right) &+ \left(\frac{H}{b_0}\right)^{2/3} \left( \frac{c_0}{b_0} x^5+\frac{b_1}{a_1} x^3z + \frac{a_2 b_0}{a_1^2} xz^2\right)\nonumber \\
	&\hspace*{-50pt}+\mathcal{O}\left(\left(\frac{H}{b_0}\right)^{4/3}\right) =0.
\end{align}
The solution of the gap equation yields the magnetic equation of state $x=x(z,h)$  in Landau theory.

The universal scaling curve of mean-field theory is obtained by taking the limit $H\to 0$ in the gap equation. In this limit only the terms grouped in the first parentheses in Eq. (\ref{eq:Landau_mEOS}) survive, while the terms proportional to $(H/b_0)^{2/3}$ provide the leading-order scaling violation. Since the latter depend on the parameters of the model, introduced in Eq. (\ref{eq:params}), they are nonuniversal and consequently model dependent.
Thus, to quantify the deviations from universal scaling,  we must specify the coefficients in the Landau effective potential. This will be done in the next section in the QM and PQM models. However, qualitative features of the scaling violation can be extracted from general considerations.
We can distinguish three asymptotic regimes:
\begin{itemize}
	\item $z\rightarrow -\infty$: In this limit  the scaling curve behaves as  $x(z)\simeq \sqrt{-z}$, and the sign of the scaling violation is determined by the sign of $(c_0/b_0 - b_1/a_1 + a_2 b_0 / a_1^2)$.
	\item $z=0$: At this point the scaling curve goes through $x=1$, and  the sign of $c_0$ determines the sign of the deviations.
	\item $z\rightarrow \infty$: In this limit  the scaling curve behaves  as $x\simeq 1/z$, and
 the sign of the correction is determined by the sign of $a_2$.
\end{itemize}
The discussion of the asymptotic behavior of the leading-order scaling violation shows that, depending the model parameters, the sign of the deviation from the scaling curve can change as a function of $z$ and can therefore cross the universal curve at several points. Indeed, Eq. (\ref{eq:Landau_mEOS}) shows that in Landau theory, the first-order correction vanishes at points $(x,z)$ where
\begin{equation}
	\frac{c_0}{b_0}+\frac{b_1}{a_1} \left(\frac{z}{x^2}\right) + \frac{a_2 b_0}{a_1^2}\left(\frac{z}{x^2}\right)^2 = 0.
\end{equation}
The roots of the quadratic equation are $z=\alpha^{\pm} x^2$,  where
\begin{equation} \label{eq:crossingalpha}
	\alpha^{\pm}= -\frac{a_1}{2a_2b_0}\left( b_1 \pm \sqrt{b_1^2-4 a_2 c_0} \right).
\end{equation}
Substitution  of  the roots into the universal curve yields the coordinates of the crossing points
\begin{equation}
	z_c^{\pm} = \frac{\alpha^{\pm}}{(1+\alpha^{\pm})^{2/3}},\quad\quad
	x_c^{\pm} = (1+\alpha^{\pm})^{-1/3}. \label{eq:crossingxz}
\end{equation}
If $\alpha^{+}$ and $\alpha^{-}$ are not real, or both of them are real and smaller than $(-1)$, then the magnetic equation of state does not cross the universal scaling function, to leading order in $H/b_0$. On the other hand, if the coefficients $\alpha^{\pm}$ are real, and only one of them is larger than $(-1)$, there is one crossing point.  Finally, if both solutions are real and larger  than $(-1)$, then there are  two crossing points.

\subsection{Quark-meson model in the mean-field approximation \label{sec:QM}}
The QM model is widely employed as a low-energy effective theory of QCD because it shares an important characteristic with QCD, namely spontaneous breaking of chiral symmetry in vacuum and  its restoration at finite temperature.  The elementary fields in this model are the quark  $q=\{u,d\}$ and meson $\phi=\{\sigma,\vec{\pi}\}$ fields, with the Lagrangian
\begin{align}
	\mathcal{L}=&\frac12 \left(\partial_{\mu}\sigma\right)^2+\frac12 \left(\partial_{\mu}\vec{\pi}\right)^2
	-U_m(\sigma,\vec{\pi}) \nonumber\\
	&+\overline{q}\left( i\gamma^{\mu}\partial_{\mu} -g\left( \sigma+i\gamma_5 \vec{\tau}\vec{\pi} \right) \right)q.
\end{align}
Here the mesonic potential is given by
\begin{equation}
	U_m(\sigma,\vec{\pi}) = \frac{\lambda}{4}\left(\sigma^2 + \vec{\pi}^2 -v^2\right)^2 - H\sigma.
\end{equation}
At low temperatures,  chiral symmetry is spontaneously broken,  and the $\sigma$ field gains a nonzero expectation value \cite{Koch:1997ei}.  In the chiral limit, $H\to 0$,  there is a   second-order phase transition at the temperature $T=T_c$. Ignoring the fluctuations of the meson fields,  the transition is governed by  mean-field dynamics, corresponding to the $O(4)$ universality class in four dimensions. In this approximation,  the dynamics of the chiral symmetry breaking can be mapped onto a Landau effective potential $\Omega(T,\mu;\sigma)$,  which is  a polynomial in the order parameter $\sigma$. The effect of vacuum and thermal fluctuations of the fermion fields are accounted for in the effective potential~\cite{Skokov:2010sf}
\begin{align}\label{qm}
	\Omega(T,&\mu;\sigma) \nonumber\\= &U_m(\sigma,0)-\frac{N_cN_f}{8\pi^2}g^4 \sigma^4 \ln\left(\frac{g\sigma}{M}\right)\nonumber\\
	&-\frac{N_cN_f}{\pi^2}\int_0^{\infty} dp p^2 T \left(
	\ln\left(1+e^{(\mu-\epsilon_q(p,\sigma))/T}\right)\right.\nonumber \\
	&+\left. \ln\left(1+e^{(-\mu-\epsilon_q(p,\sigma))/T}\right)\right),
\end{align}
where $\epsilon_q(p,\sigma)=\sqrt{p^2+g^2\sigma^2}$. The values of the parameters $\lambda$,  $v^2$ and $H$ are  determined by choosing the pion mass $m_{\pi}$, the sigma mass $m_{\sigma}$, as well as the pion decay constant $f_{\pi}$ in vacuum. The Yukawa coupling $g$ is set by the constituent quark mass in vacuum. In the chiral limit, $H\to 0$, the pion is a true Goldstone boson, with a vanishing vacuum mass. The parameter $M$ is an arbitrary renormalization scale. Modifications of $M$ can be absorbed by redefining $\lambda$ and $v^2$.

As pointed out in Ref.\ \cite{Skokov:2010sf},  both the vacuum and thermal contributions contain a nonanalytic term in $\sigma$,  which cancel at nonzero temperature. This cancellation is crucial for obtaining a second-order chiral transition in the chiral limit. Close to the critical point, where  $H \approx 0$ and $T\approx T_c$,  the order parameter  $\sigma$ is very small. Consequently, the  contribution of  the thermal fermion loop to the Landau effective potential can be obtained in the high-temperature expansion \cite{Dolan:1973qd,Klajn:2013gxa}. We thus obtain the Landau free energy density
\begin{equation}
	\Omega(T,H;\sigma) =  a(t)\frac{\sigma^2}{2}+
	b(t)\frac{\sigma^4}{4}+
	c(t)\frac{\sigma^6}{6} + \dots -H\sigma,
\end{equation}
where  the coefficients are functions of the reduced temperature $t$ and the input parameters
\begin{align}\label{eq:HTEcoeffs}
\begin{split}
	a(t)&=\left( \frac{N_cN_f g^2 T_c^2 (2t+t^2)}{6} \right),\\
	b(t)&= \frac{m_{\sigma}^2-m_{\pi}^2}{2 f_{\pi}^2} + \frac{N_cN_f g^4}{2\pi^2} \left(\gamma_E-\ln\left(\frac{\pi T_c (1+t)}{g f_{\pi}} \right)\right)
	,\\
	c(t)&= -\frac{7 \zeta(3) N_cN_f g^6}{32 \pi^4 T_c^2(1+t)^2}.
\end{split}
\end{align}
In the mean-field approximation,  the second-order chiral phase transition appears at
\begin{equation}
	T_c = \sqrt{\frac{3\left( m_{\sigma}^2-3m_{\pi}^2 \right)}{N_cN_f g^2} +\frac{3g^2 f_{\pi}^2}{2\pi^2}}.
\end{equation}

In this approximation, the QM model is a particular realization of Landau theory. Using the coefficients of the effective potential, Eq. (\ref{eq:HTEcoeffs}), we can, following the discussion in the previous section,
compute $h_0$, $t_0$ and $z_0$ and explicitly determine the magnetic equation  of state given in Eq. (\ref{eq:Landau_mEOS}), including the leading-order scaling violating term.

\subsection{Polyakov loop extended quark-meson model \label{sec:PQM}}

The chiral QM model is an effective
realization of the chiral  sector of  QCD. However, because the local
$SU(N_c)$ invariance of QCD is replaced by a global symmetry in the model, color confinement is lost. Nevertheless, the confining properties
of QCD can be approximately accounted for by including the expectation value of the Polyakov loop
\begin{equation}
\Phi =            \frac{1}{N_c}\left\langle \mathrm{Tr}_c\, L(\vec{x}) \right\rangle, \quad
\overline{\Phi} = \frac{1}{N_c}\left\langle \mathrm{Tr}_c\, L^{\dagger}(\vec{x}) \right\rangle ,
\end{equation}
with
\begin{equation}\label{loop}
	L(\vec{x}) = \mathcal{P} \exp\left( i\int_0^{\beta} d\tau A_4(\vec{x},\tau) \right),
\end{equation}
in a low-energy chiral effective model, like the QM model
\cite{Fukushima:2002ew,Fukushima:2003fm,Fukushima:2003fw,Schaefer:2007pw,Skokov:2010wb,Skokov:2010uh}.
Here $A_4=iA_0$  is the temporal component of the Euclidean gluon field, $\beta = 1/T$ and $\mathcal{P}$ denotes path ordering.
Thus, the PQM model effectively combines both the chiral symmetry and confinement of QCD.

The Lagrangian of the PQM model reads
\begin{align}
	\mathcal{L} = &\overline{q}\left( i\gamma^{\mu} D_{\mu} -g(\sigma +i\gamma_5 \vec{\tau}\vec{\pi})\right)q \nonumber \\ &+\frac12 \left(\partial_{\mu}\sigma\right)^2+\frac12 \left(\partial_{\mu}\vec{\pi}\right)^2
	-U_m(\sigma,\vec{\pi})-\mathcal{U}(\Phi,\overline{\Phi}).
\end{align}
The coupling between the effective
gluon field and quarks is implemented through the covariant derivative,
$D_{\mu}=\partial_{\mu}+iA_{\mu}$,
where  the spatial components of the gluon
field are neglected, i.e. $A_{\mu}=\delta_{\mu0}A_0$.
Here $\mathcal{U}(\Phi,\overline{\Phi})$ is the potential for the thermal expectation value of the Polyakov loop.

The thermodynamic potential in the PQM model is given by \cite{Skokov:2010sf}
\begin{align} \label{eq:PQMOmega}
	\Omega(\sigma,\Phi,\overline{\Phi};T,\mu) = &U_m(\sigma,0)
	-\frac{N_cN_f}{8\pi^2}g^4 \sigma^4 \ln\left(\frac{g\sigma}{M}\right) \nonumber\\
	&+\Omega_f^{th} (T,\mu;\sigma,\Phi,\overline{\Phi}) +\mathcal{U}(\Phi,\overline{\Phi};T)
\end{align}
where the meson and vacuum fermion contributions are the same as in the QM model in Eq. (\ref{qm}), whereas the thermal fermionic contribution is modified due to coupling of quarks to the Polyakov loop background
\begin{align}\label{fermion}
	\Omega_f^{th} (T,&\mu;\sigma,\Phi,\overline{\Phi}) \nonumber \\= &-\frac{N_f T}{\pi^2}
	\int_0^{\infty} dp p^2 \bigg( \ln g^{(+)}(T,\mu;\sigma,\Phi,\overline{\Phi}) \nonumber \\ &+\ln g^{(-)}(T,\mu;\sigma,\Phi,\overline{\Phi}) \bigg)
\end{align}
with
\begin{align}
	g^{(+)}(T,\mu;\sigma,\Phi,\overline{\Phi}) = &1 + 3\Phi e^{-(E_q-\mu)/T}+ 3\overline{\Phi} e^{-2(E_q-\mu)/T} \nonumber \\
	&+ e^{-3(E_q-\mu)/T},\nonumber\\
	g^{(-)}(T,\mu;\sigma,\Phi,\overline{\Phi}) = &g^{(+)}(T,-\mu;\sigma,\overline{\Phi},\Phi).
\end{align}
Clearly, by taking the limit  $\Phi\to 1$ in Eq. (\ref{fermion}), one recovers the fermion part of the effective potential of the QM model, Eq. (\ref{qm}).
The gluon potential, $\mathcal{U}(\Phi,\overline{\Phi};T)$, is constructed so as to respect the $Z(N_c)$ global symmetry,  with parameters chosen to reproduce the thermodynamics of pure lattice gauge theory~\cite{Ratti:2005jh,Roessner:2006xn,Lo:2013hla}. We use the potential obtained in Ref.~\cite{Roessner:2006xn}
\begin{equation}
	\frac{\mathcal{U}(\Phi,\overline{\Phi};T)}{T^4} = -\frac12 a(T)\left( \Phi\overline{\Phi} \right)
	+ b(T) \log M_H(\Phi,\overline{\Phi})
\end{equation}
with
\begin{equation}
	M_H(\Phi,\overline{\Phi})=1-6\Phi\overline{\Phi}
	+ 4\left( \Phi^3 + \overline{\Phi}^3 \right) -3\left(\Phi\overline{\Phi}\right)^2.
\end{equation}
The temperature-dependent coefficients are given by
\begin{equation}
	a(T)=a_0 + a_1 \left(\frac{T_0}{T}\right) + a_2 \left(\frac{T_0}{T}\right)^2,\quad
	b(T)= b_3 \left(\frac{T_0}{T}\right)^3
\end{equation}
with the parameters
\begin{align}
a_0 &=3.51,\quad\quad a_1=-2.47,\quad\quad a_2=15.2,\phantom{xxxxxxxxxx}\nonumber\\ b_3 &=-1.75,\quad\quad T_0=270\; \mathrm{MeV}.
\end{align}

In the mean-field approximation, the  expectation value of $\sigma$ and of the Polyakov loop,  $\Phi$ and $\overline{\Phi}$ are determined
by requiring that the thermodynamic potential is~\footnote{The stationary point is a saddle point in the variables $\Phi$ and $\bar{\Phi}$. Nevertheless, for the effective Polyakov loop potential employed in this paper, the system is thermodynamically stable, as shown in ref.~\cite{Sasaki:2006ww}.}
\begin{equation}
	\frac{\partial \Omega}{\partial \sigma} = \frac{\partial \Omega}{\partial \Phi} = \frac{\partial \Omega}{\partial \overline{\Phi}}=0.
\end{equation}
The model parameters are fixed by requiring that the
vacuum physics is reproduced, as indicated in Sec.~\ref{sec:QM} for the QM model.

In the chiral limit,  this  model  exhibits  second-order chiral phase transition with mean-field exponents. Near the critical point, the thermodynamic potential is a polynomial in the order parameter $\sigma$, as in Eq. (\ref{landau}), with  coefficients that can be extracted from  Eq.  (\ref{eq:PQMOmega}) using the  high-temperature expansion  \cite{Klajn:2013gxa}. In this case, however, the coefficients of the potential  depend on the expectation value of the Polyakov-loop and cannot be obtained in a closed form. Thus, for the PQM model, the critical temperature, the coefficients of the  Landau potential and the magnetic equation of state are computed numerically.

\subsection{\textit{O(N)} linear sigma model in the  large-\textit{N} limit  \label{sec:Oinfintro}}
In the preceding sections we have introduced the Landau effective action and two chiral effective models, which allow us to explore various aspects of the magnetic equation of state in the  mean-field approximation. In this section we turn to the $O(N)$ symmetric linear sigma model,  where the thermodynamic potential and the scaling properties near the critical point can be computed exactly in the  large-$N$ limit.  The LS model in (3+1) dimensions is described by the Euclidean action

\begin{align}\label{ns}
	S_E = &\int d^4 x\left( \frac{1}{2}(\partial_{\mu}\sigma)^2 + \frac{1}{2}(\partial_{\mu}\pi_i)^2 + \frac{1}{2}m^2(\sigma^2+\pi_i^2 ) \right.\nonumber \\
	&+\left.\frac{\lambda}{N}\left(\sigma^2+\pi_i^2\right)^2-\frac{\sqrt{N}}{2}H \sigma\right),
\end{align}
where  the subscript $\mu$ denotes a direction in Euclidean space-time while $i$ is an index in flavor space, spanned by the $N$-component vectors $\lbrace\sigma,\pi_i\rbrace$. The $N$-dependent factors in Eq. (\ref{ns}) are introduced for later convenience.

For  vanishing external field $H$,  the action is invariant under rotations in the $N$-dimensional flavor space.  For negative values of $m^2$, this symmetry is spontaneously broken in the vacuum and the $N$-tuple $\lbrace\sigma,\pi_i\rbrace$ acquires a  nonzero expectation value, a condensate. The coordinates in flavor space are chosen such that the condensate is in the $\sigma$ direction. Consequently, the $N-1$ remaining fields, $\pi_i$, have a vanishing expectation value. The condensate $\langle\sigma\rangle$ is an order parameter of the spontaneously broken $O(N)$ symmetry and the shifted field $\sigma^\prime=\sigma-\langle\sigma\rangle$ represents the fluctuations of the $\sigma$ field about its expectation value.

To determine the thermodynamic potential density $\omega[T]$,  we  employ  the $2PI$ formalism \cite{Baym:1962sx,Cornwall:1974vz}. The $2PI$ functional for the theory yields
\begin{align}
	\omega[T;\langle &\sigma \rangle, G_{\pi},G_{\sigma}] \nonumber\\
	=
	&\frac{T}{V} \biggl( S_E[\langle \sigma \rangle]  +
	\frac{N-1}{2}\mathrm{Tr}\ln G_{\pi}^{-1} \nonumber\\ 
	&+ \frac{N-1}{2}\mathrm{Tr}\left( \left(D_{\pi}^{-1}-G_{\pi}^{-1}\right)G_{\pi} \right) + \frac{1}{2}\mathrm{Tr}\ln G_{\sigma}^{-1} \nonumber \\
	&+\left.\frac{1}{2}\mathrm{Tr}\left( \left(D_{\sigma}^{-1}-G_{\sigma}^{-1}\right)G_{\sigma} \right)
	  -\Gamma_2[\langle \sigma \rangle, G_{\pi},G_{\sigma}] \right),\label{scalar}
\end{align}
where  $D_{\sigma}$, $D_{\pi}$, $G_{\sigma}$ and $G_{\pi}$ are the bare and dressed propagators of the sigma and pion fields, respectively. Moreover, $\Gamma_2[\langle \sigma \rangle, G_{\pi},G_{\sigma}]$ denotes the sum of all possible $2PI$ diagrams (with dressed propagators), while  $V$ and $T$ are the volume and the temperature of the system. Finally, the trace in Eq. (\ref{scalar}) is given by
\begin{equation}
\mathrm{Tr}=T\sum_n\int\frac{d^3k}{(2\pi)^3},
\end{equation}
where the sum is over the Matsubara frequencies.

The physical values of the dressed propagator and the expectation value of the field are determined  by the stationarity conditions
\begin{equation}
	\frac{\delta\omega}{\delta \langle \sigma \rangle} =0,\quad\quad
	\frac{\delta\omega}{\delta G_{\pi}}=0,\quad\quad
	\frac{\delta\omega}{\delta G_{\sigma}}=0.
	\label{eq:LargeNstationarity}
\end{equation}
The second and the third equations are just Dyson equations for the pion and sigma fields respectively,
\begin{equation}
G^{-1}=D^{-1}-2\, \frac{\delta \Gamma_2}{\delta G}=D^{-1}+\Sigma,
\end{equation}
where $-2\, \delta \Gamma_2/\delta G$ is identified with the self-energy $\Sigma$.
A tractable self-consistent scheme for calculating the thermodynamic potential starting from Eq. (\ref{scalar}) is defined by a choice of the set of $2PI$ diagrams contributing to $\Gamma_2$ (for details see Ref.~\cite{Baym:1962sx}).

For convenience we simplify our notation by introducing $\phi \equiv 2\langle \sigma \rangle /\sqrt{N}$. With this choice the inverse bare Euclidean propagators are given by
\begin{align}
	D^{-1}_{\sigma}(\mathbf{k},n) &= \mathbf{k}^2 + \omega_n^2 + m^2 + 3\lambda \phi^2,\\
	D^{-1}_{\pi}(\mathbf{k},n) &= \mathbf{k}^2 + \omega_n^2 + m^2 + \lambda \phi^2,
\end{align}
where $\omega_n = 2n \pi T$ is the Matsubara frequency and $\mathbf{k}$ denotes the momentum in the spatial direction. The bare mass-squares of the $\sigma$ and  $\pi$ fields are given by $m^2+3\lambda \phi^2$ and $m^2+\lambda \phi^2$, respectively.

\begin{figure}[tb]
\centering
   \includegraphics[width=0.4\linewidth]{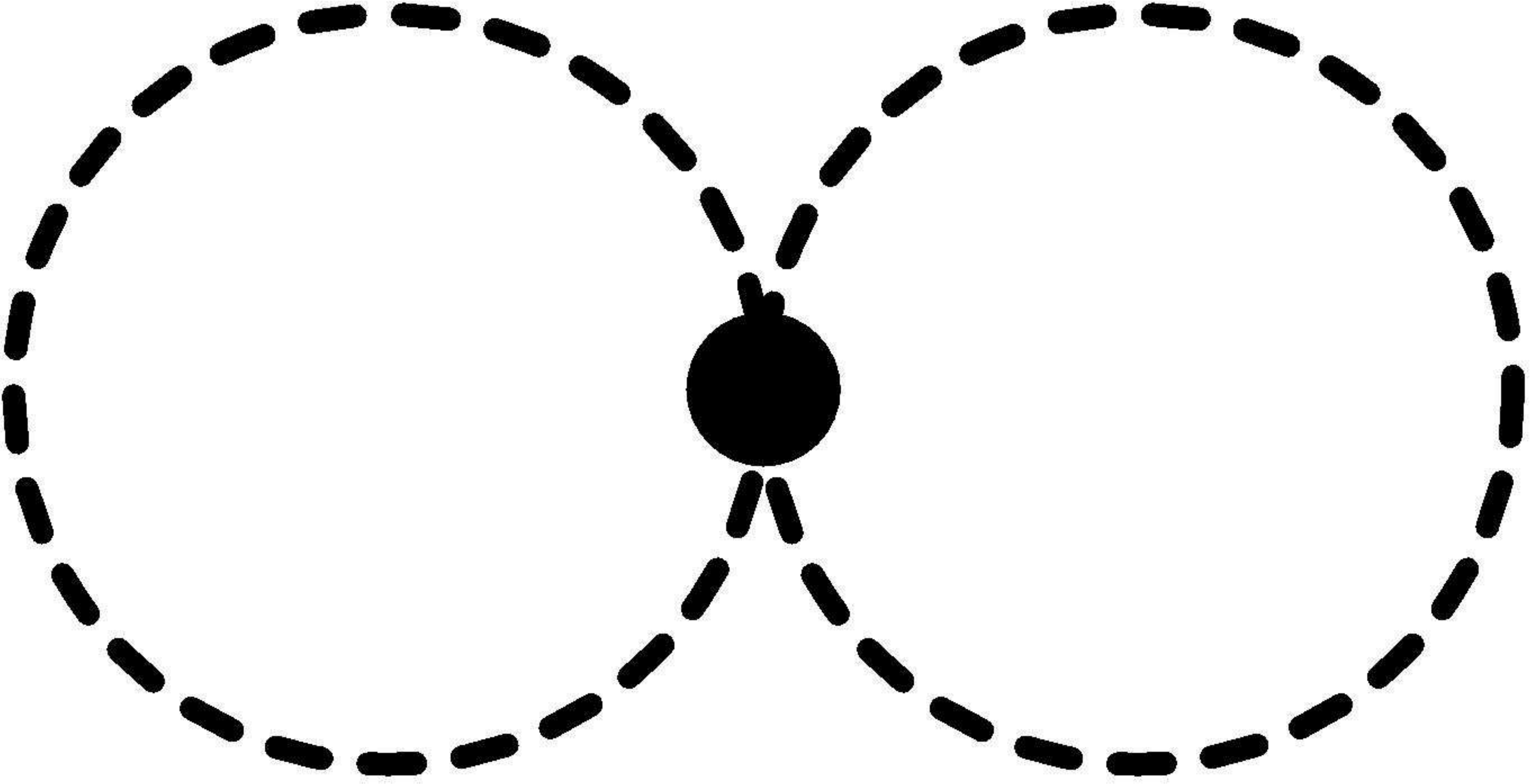}
   \caption{ The  only contribution to $\Gamma_2$  in the $1/N$ expansion  of the  $O(N)$ sigma model for  $N\to\infty$.   The dashed lines are pion propagators, whereas  the filled dot depicts  the four-point vertex, with coupling strength $\lambda / N$. \label{fig:diagram}}
\end{figure}

In the $O(N)$  linear sigma model,  at $N\to \infty$, the contributions of sigma loops  to $\Gamma_2[\langle \sigma \rangle, G_{\pi},G_{\sigma}]$ are suppressed,  due to the $1/N$ factor introduced in the four-point coupling in the action. Consequently, to leading order in $1/N$,  the only relevant contribution to the $2PI$ diagrams is the two-pion loop diagram shown in Fig. \ref{fig:diagram},
\begin{equation}
\Gamma_2[\phi,G_{\pi}]=-N\lambda(\mathrm{Tr}\,G_{\pi})^2.
\end{equation}
This in turn yields the pion self-energy
\begin{equation}
	\Sigma_{\pi}(\mathbf{k},n) = 4\lambda \mathrm{Tr}\,G_{\pi}.
\end{equation}
Since the fluctuations of the $\sigma$ field are neglected in the large-$N$ limit,  the  thermodynamic potential density depends only on the condensate $\langle\sigma\rangle$ and on the dressed pion propagator $G_\pi$. These are then determined by the first two equations of Eq. (\ref{eq:LargeNstationarity}). Up to leading order in $N$ the potential reads
\begin{align}\label{2PI:vector}
	\omega[T;\phi,G_{\pi}] =
	&U[\phi] +
	 \frac{N}{2}\mathrm{Tr}\ln G_{\pi}^{-1} \nonumber \\
	 &+ \frac{N}{2}\mathrm{Tr}\left( \left(D_{\pi}^{-1}-G_{\pi}^{-1}\right)G_{\pi} \right) \nonumber \\
	 &+N\lambda \left(\mathrm{Tr} G_{\pi}\right)^2,
\end{align}
with
\begin{equation}
	U[\phi] = \frac{N}{4}\left(\frac{m^2}{2}\phi^2 + \frac{\lambda}{4}\phi^4 - H\phi\right).
\end{equation}
The Dyson equation in this approximation yields
\begin{align}
	G^{-1}_{\pi}(\mathbf{k},n) &= D^{-1}_{\pi}(\mathbf{k},n) + \Sigma_{\pi}(\mathbf{k},n) \nonumber\\ &=
	\mathbf{k}^2 + \omega_n^2 + m^2 + \lambda \phi^2 + 4\lambda \mathrm{Tr}G_{\pi}.
\end{align}
This is a self-consistent equation for the self-energy or equivalently for the renormalized pion mass. This is readily seen by rewriting the propagator in the compact form
\begin{equation}
	G_{\pi}(\mathbf{k},n) = \left( \mathbf{k}^2 + \omega_n^2 + M_{\pi}^2 \right)^{-1},
\end{equation}
where the pion mass $M_{\pi}$ is a solution of  the equation
\begin{equation}
	M_{\pi}^2 = m^2 + \lambda \phi^2 + 4\lambda \mathrm{Tr}G_{\pi} (M_{\pi}).
\end{equation}
The boson loop integral $\mathrm{Tr}\,G_{\pi}$  can conveniently be expressed in terms of the logarithmic term in Eq. (\ref{2PI:vector}),
\begin{align}
\label{gg}
	\mathrm{Tr}(G_{\pi}) &= \frac{\partial}{\partial M_{\pi}^2}  \mathrm{Tr}\ln(G_{\pi}^{-1}). \nonumber\\
	&= \frac{\partial}{\partial M_{\pi}^2} T\sum_{n}\int \frac{d^3 k}{(2\pi)^3} \ln\left(k^2 + (2n\pi T)^2 +M_{\pi}^2\right)\nonumber\\
	&=  \frac{\partial}{\partial M_{\pi}^2} \int \frac{d^3 k}{(2\pi)^3} \left(\sqrt{k^2+M_{\pi}^2} \right.\nonumber\\ &\phantom{=}+\left. 2T \ln\left( 1-e^{-\sqrt{k^2+M_{\pi}^2}/T} \right)\right).
\end{align}
The first term is the UV divergent vacuum contribution, while the second term is the finite temperature contribution. Using dimensional regularization,  we retain only the finite part of the vacuum integral, following \cite{vanHees:2001ik,VanHees:2001pf,vanHees:2002bv,Lenaghan:1999si}. The renormalized vacuum contribution is then given by
\begin{equation}
	 \int \frac{d^3 k}{(2\pi)^3} \frac{1}{2\sqrt{k^2+M_{\pi}^2}} \rightarrow
	 \frac{1}{(4\pi)^2}\left( M_{\pi}^2 \ln \frac{M_{\pi}^2}{\mu^2} -M_{\pi}^2 + \mu^2 \right),
\end{equation}
where $\mu$ is an arbitrary renormalization scale. We  choose $\mu = m_{\pi}=138\, \textrm{MeV}$, which simplifies the formulas somewhat.

Stationarity of the functional given in Eq. (\ref{2PI:vector}) hence yields the following system of equations for the renormalized pion mass $M_{\pi}$ and the order parameter $\phi$:
\begin{align}
\begin{split}
	\label{eq:Oinfeq}
	H &= M_{\pi}^2\phi, \\
	M_{\pi}^2 &= m^2 + \lambda \phi^2 + 4\lambda \mathrm{Tr}G_{\pi}.
\end{split}
\end{align}
The model parameters $\lambda$, $m^2$ and $H$ are chosen so as to reproduce the vacuum pion and sigma mass,  as well as  the pion decay constant. These conditions yield the following constraints \cite{Lenaghan:1999si}:
\begin{align}
	H&=m_{\pi}^2 f_{\pi},\quad\quad \lambda = \frac{m_{\sigma}^2 - m_{\pi}^2}{2 f_{\pi}^2}, \nonumber\\
	m^2 &= -\frac{m_{\sigma}^2-3 m_{\pi}^2}{2}
	-\frac{\lambda}{4\pi^2}\left( m_{\pi}^2 \ln\frac{m_{\pi}^2}{\mu^2}-m_{\pi}^2 + \mu^2 \right) \nonumber\\
	&=-\frac{m_{\sigma}^2-3 m_{\pi}^2}{2},
\end{align}
where in the last equality we used $\mu=m_\pi$.
To derive the magnetic equation of state for this model, we again apply the high temperature expansion \cite{Dolan:1973qd}.

\subsubsection{The magnetic equation of state of the LS model}

Near the critical point,  i.e. where   $H\approx 0$,  $\phi\approx 0$ and  the pion mass  $M_\pi\approx 0$,  we can expand $\mathrm{Tr}(G_{\pi})$ in powers of $M_\pi/T$ by using  the high-temperature expansion,
\begin{align}\label{expansion}
	\mathrm{Tr}(G_{\pi})
	&= \left(\frac{T^2}{12}+\frac{\mu^2}{16\pi^2}\right)
	-\frac{T}{4\pi}M_{\pi} \nonumber \\ &\phantom{=}+
	\frac{\ln\left(\frac{4\pi T}{\mu}\right)-\gamma_E}{8\pi^2} M_{\pi}^2 + \mathcal{O}\left(M_{\pi}^4\right).
\end{align}
Here $\gamma_E$ is the Euler-Mascheroni constant.

By substituting the leading term in Eq.  (\ref{expansion})  into  the gap equation, Eq. (\ref{eq:Oinfeq}), with $M_\pi=\phi=0$, one  finds  the critical temperature for the second-order transition~\cite{Lenaghan:1999si}
\begin{equation}
	m^2 + 4\lambda \left( \frac{T_c^2}{12}+\frac{\mu^2}{16\pi^2}\right) = 0 \rightarrow
	T_c = \sqrt{3}\,f_{\pi}^{ch},
\end{equation}
where
\begin{equation}\label{eq:fpi}
f_{\pi}^{ch}=  f_{\pi}\left(\frac{m_{\sigma}^2-3 m_{\pi}^2}{m_{\sigma}^2-m_{\pi}^2}-\frac{1}{4\pi^2}\frac{m_{\pi}^2}{f_{\pi}^2}\right)^{1/2}
\end{equation}
is the pion decay constant in the chiral limit.
In Eq. (\ref{eq:fpi}) we have inserted the renormalization scale $\mu = m_{\pi}$. Using Eq.~(\ref{eq:Oinfeq}) we also find  that  for  $H=0$ and  $M_{\pi}=0$  the order parameter in the broken phase is given by
\begin{equation}
	\label{eq:phiT}
	\phi = \sqrt{\frac{T_c^2-T^2}{3}}.
\end{equation}
Thus, near the critical point the order parameter scales as  $\phi \sim (T_c-T)^{1/2} \sim (-t)^{1/2}$, with the critical exponent $\beta=1/2$.

In order to obtain the exponent $\delta$ and the magnetic equation of state,  we retain only the leading (linear) term in $M_\pi$ in the second  equation in Eq.~(\ref{eq:Oinfeq}). This leads  to the system of equations
\begin{equation}
\begin{split}
	\label{eq:Oinfeq2}
	H &= M_{\pi}^2\phi, \\
	0 &= \lambda \left(\frac{T_c^2-T^2}{3} - \phi^2 +\frac{T}{\pi}M_{\pi}\right).
\end{split}
\end{equation}
Consequently, at $T=T_c$, we find
\begin{equation}
	\phi \sim H^{1/5}, ~~~{\rm and } ~~~~M_{\pi} \sim H^{2/5}.
\end{equation}
Thus,  the critical exponent $\delta = 5$, as in the spherical model in three spatial dimension~\cite{zinn2002quantum, Baxter:1982zz}.
We note that in four dimensions, the model yields $\beta=1/2$ and $\delta=3$, as in the mean-field case.

We are now ready to derive the magnetic equation of state, including the leading-order scaling violating term. By eliminating the pion mass in Eq.~(\ref{eq:Oinfeq}) and using the high-temperature expansion of the one-loop self-energy given in Eq.~(\ref{expansion}), one arrives at the gap equation
\begin{equation}\label{new-gap}
\frac{T_c}{\pi}(1+t)\left(\frac{H}{\phi}\right)^{1/2}=\phi^2+\frac{ T_c^2}{3}(2t+t^2)+\frac{3\,\alpha}{4\,\pi^2}\frac{H}{\phi},
\end{equation}
which is valid near the critical point, where $t$, $H$ and $\phi$ are small. Here we introduced the shorthand notation
\begin{equation}
\alpha=\frac{2}{3}\left[\ln\left(\frac{4\pi T_c}{\mu}\right)-\gamma_E\right]-\frac{4\,\pi^2}{3\,\lambda}
\end{equation}
and neglected the temperature dependence of the logarithm, which yields only terms of higher order in the scaling violating field.
In analogy with Eqs.~(\ref{eq:scal-var}) and (\ref{eq_Landau_xzdef}), we introduce the scaling fields $z,x$ and $h$
by means of
\begin{equation}
H=h\,h_0\,H_0,\qquad t=z\, t_0\,h^{2/5},\qquad\phi=x\,\phi_0\,h^{1/5}.
\end{equation}
The constants $h_0$ and $t_0$ are determined by the  normalization conditions
\begin{equation}
x(z=0)=1,\qquad\lim_{z\to -\infty} x(z)/(-z)^{\beta}=1,
\end{equation}
which are equivalent to Eq. (\ref{eq14}). One finds
\begin{equation}
h_0=\frac{\phi_0^5\,\pi^2}{H_0\,T_c^2},\qquad t_0=\frac{3\,\phi_0^2}{2\,T_c^2},
\label{eq:Oinf_t0h0}
\end{equation}
which depend explicitly on the normalization scale $\phi_0$, while the ratio
\begin{equation}
z_0 = \frac{h_0^{2/5}}{t_0} = \frac{2}{3} \left(\frac{\pi^4 T_c^6}{H_0^2}\right)^{1/5}
\end{equation}
depends, as expected, only on $H_0$.

The gap equation, expressed in terms of the scaling variables, is now obtained by squaring Eq.~(\ref{new-gap}) and consistently retaining terms up to order $h^{2/5}$,
\begin{equation}\label{ON-scaling-eos}
\left[x(x^2+z)^2-1\right]+h^{2/5}\,t_0 \left[(x^5-1+\alpha)(x^2+z)\right]=0.
\end{equation}
In the limit $h\to 0$, only the first term in square brackets in Eq. (\ref{ON-scaling-eos}) survives. This yields the universal scaling function for the $O(N)$ linear sigma model in the $N\to \infty$ limit.
More generally, for nonzero $h$, the solution of Eq. (\ref{ON-scaling-eos}) yields the magnetic equation of state, including the leading-order scaling violation.

The  subleading term in Eq.~(\ref{ON-scaling-eos}) is not unique, since it may be modified by using the leading-order (scaling) magnetic equation of state. The form given here was obtained by eliminating terms with noninteger powers of $x$ as well as those involving higher powers than linear in $z$. Another form of this term leads to a modified magnetic equation of state for nonzero $h$. However, the difference is of higher order, i.e. at least of order $h^{4/5}$. Clearly, other forms of the leading-order scaling violating term in Eq.~(\ref{eq:Landau_mEOS}) can be obtained in an analogous manner. We note that the nonuniqueness of the leading-order symmetry breaking term does not affect the location of the possible crossing points, discussed in Sec.~\ref{sec:Landau-theory}.

\section{Model dependence of  scaling properties of the order parameter   \label{sec:mEoS}}

\begin{figure}[bt]
	\centering
	\includegraphics[width=0.95\linewidth]{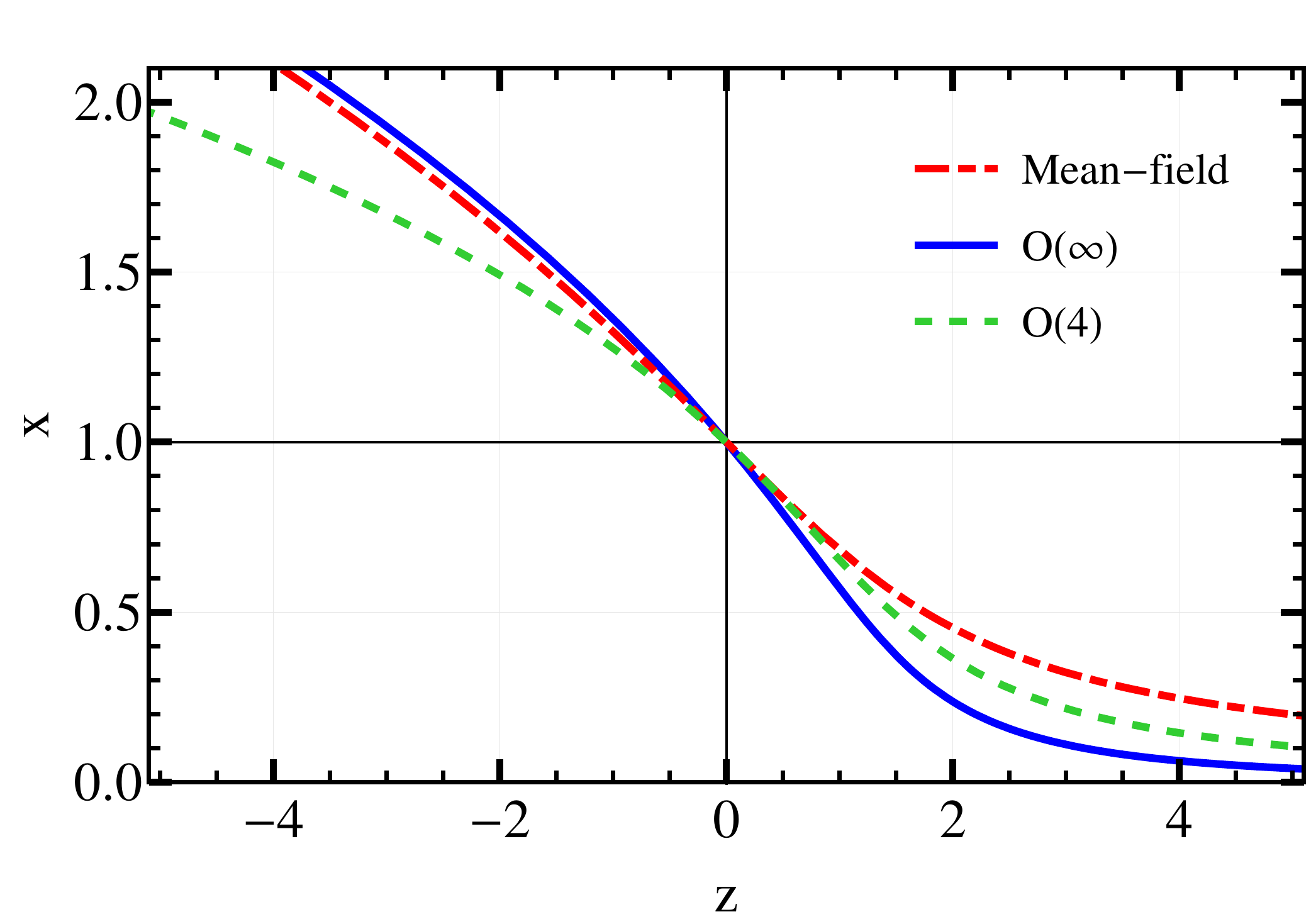}
	\caption{ The scaling function of the order parameter in   mean-field models,   the  $O(N)$ linear sigma model in the $N\to\infty$ limit, and in the $O(4)$ universality class.  \label{fig:scaling_curves}}
\end{figure}

\begin{figure*}[t]
	\includegraphics[width=0.49\linewidth]{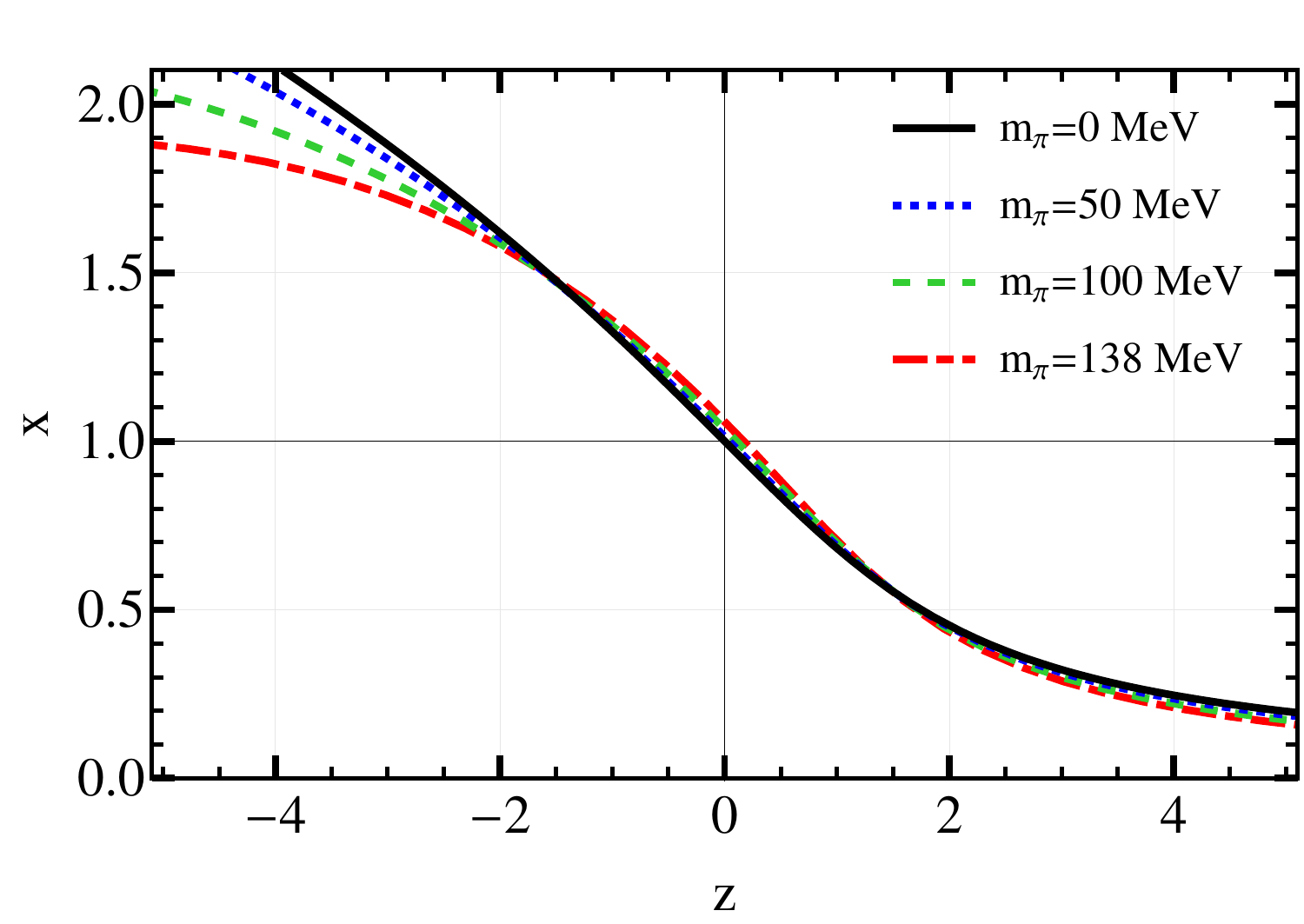}
	\includegraphics[width=0.49\linewidth]{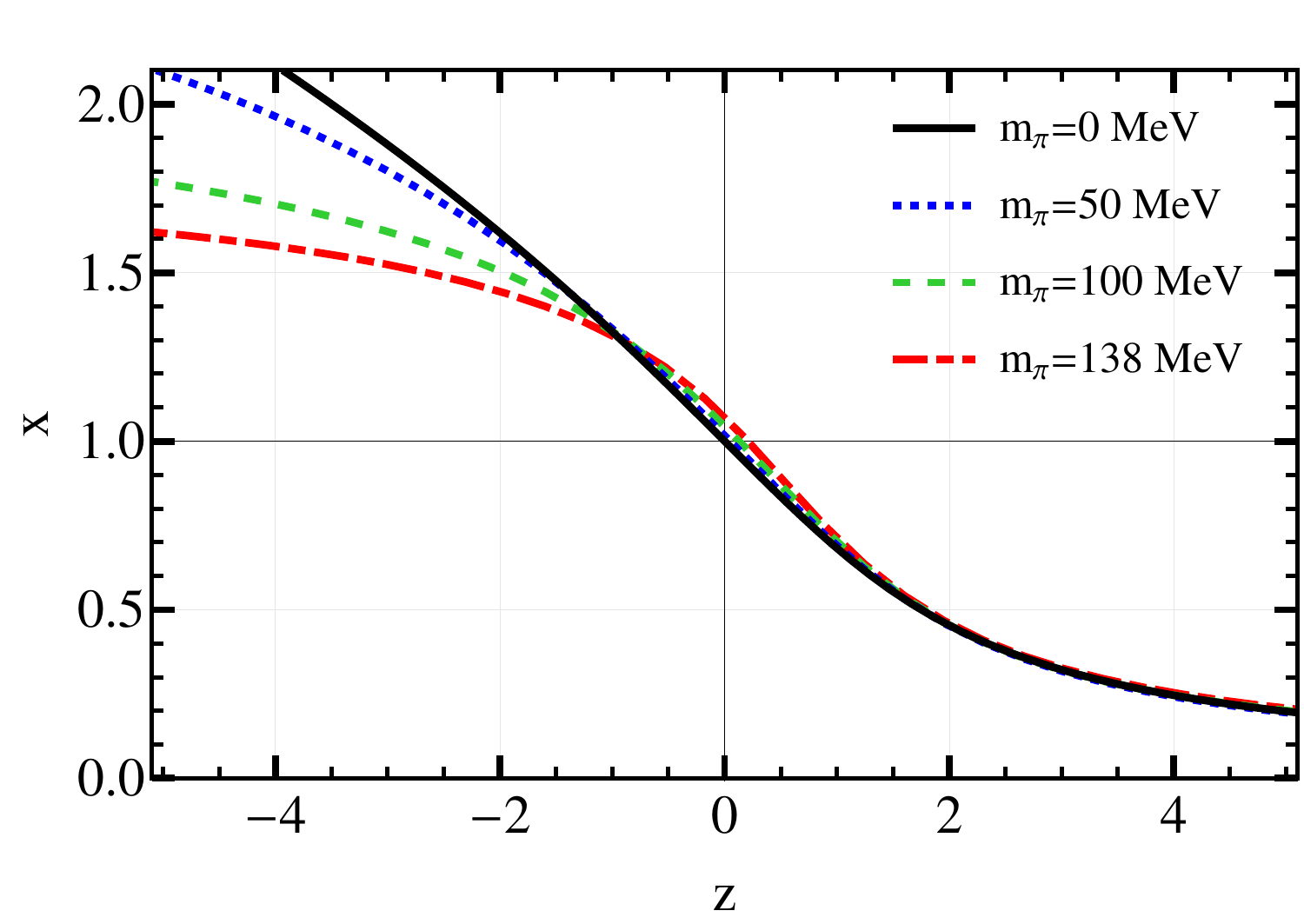}
	\caption{The magnetic equation of state for the QM (left) and PQM (right) models. The black line  corresponds to the universal scaling curve, which coincides in these two models. In both models the sigma mass was fixed to $m_{\sigma}=400\; \mathrm{MeV}$, whereas the constituent quark mass in the vacuum was set to $m_{q}=300\; \mathrm{MeV}$.\label{fig:QMPQMmanycurves}}
\end{figure*}

In the preceding section we have computed the magnetic equation of state in three different models. The QM model and its Polyakov loop extended counterpart, the PQM model, were both evaluated in the mean-field approximation. Consequently,  the corresponding scaling functions coincide and are given by the solution of the gap equation
\begin{equation}\label{ee1}
	x(x^2 + z) = 1.
\end{equation}
The corresponding equation in the $O(N)$ linear sigma model in the large-$N$ limit differs from Eq. (\ref{ee1}), and reads
\begin{equation}\label{ee2}
	x(x^2 + z)^2 = 1.
\end{equation}

In Fig. \ref{fig:scaling_curves} we show the scaling functions,
given by the solutions  $x=f_G(z)$  of Eqs.  (\ref{ee1}) and (\ref{ee2}).
In the broken phase the universal curves are close to each other, while  in the restored phase, they differ considerably.
On a qualitative level, this behavior can be understood by considering the structure of the magnetic equation of state in the asymptotic regions  $z\to \pm \infty$.  For large negative $z$, the scaling function $f_G(z)$ is of the form $(-z)^{\beta}$, whereas  for positive $z$ it asymptotically approaches $z^{-\gamma}$. Since in both models the   critical exponent $\beta=1/2$,
the two universal curves are very similar for $z<0$. On the other hand,
$\gamma=1$ in the mean-field QM and PQM models and $\gamma=2$ in the large-$N$ linear sigma model. This difference is clearly reflected in the magnetic equation of state in the restored phase, i.e. for $z>0$.

For  comparison, the scaling equation of state of the $O(4)$ universality class, obtained in lattice simulations \cite{Engels:2014bra}, is  also shown. There are clear differences between the model results and the $O(4)$ universality class. Again, the characteristics can be understood in terms of the values of the $\gamma$ and  $\beta$ exponents.

Since the QM and PQM models belong to the $O(4)$ universality class \cite{Bohr:2000gp}, differences between the scaling properties of these models  and the $O(4)$ universality class, seen in Fig. \ref{fig:scaling_curves}, will disappear when the effect of fluctuations is properly included in the thermodynamic potential.

Recent LQCD  studies~\cite{Ding:2015pmg} of the chiral phase transition with (2+1) flavors indicate that the scaling violation seen in the QCD magnetic equation of state remains moderate up to physical values of the light quarks masses. Moreover, the nonuniversal parameters $h_0$ and $t_0$ for QCD were determined. In this section we assess the scaling violation in the models presented above, and compare the nonuniversal parameters extracted in the models with the lattice QCD results. The numerical results presented in this section are based on the full thermodynamic potentials (Eqs. \eqref{qm}, \eqref{eq:PQMOmega} and \eqref{2PI:vector}) without invoking the high-temperature expansion.

\begin{figure}[b]
	\centering
	\includegraphics[width=0.95\linewidth]{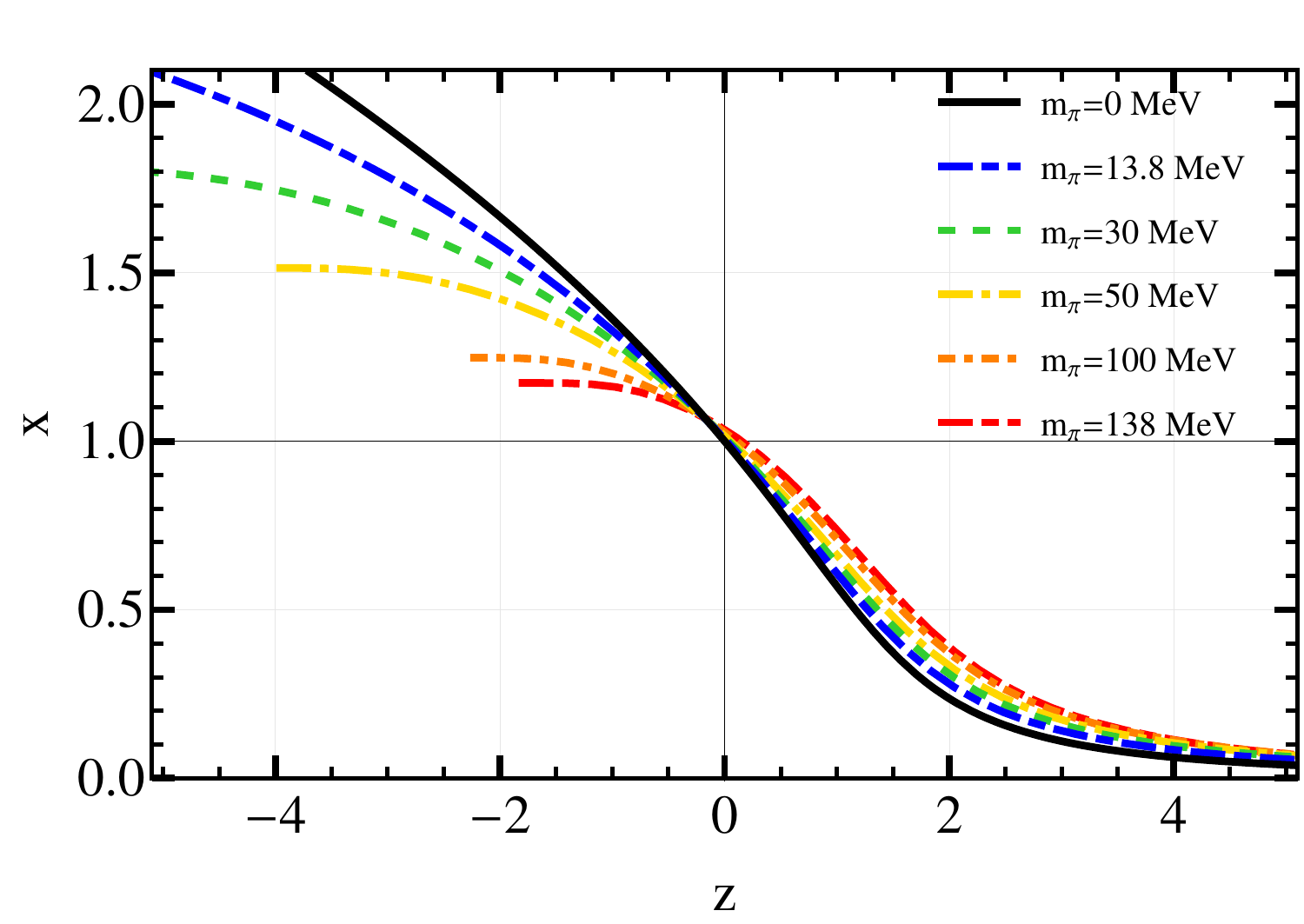}
	\caption{The magnetic equation of state for the $O(N)$ linear sigma model in the  $N\to\infty$ limit. The black line  corresponds to the universal scaling curve of the $O(N=\infty)$ universality class. The end points of the curves at negative $z$ values correspond to $T=0$. In  this calculation,  $m_{\sigma}=400\; \mathrm{MeV}$.\label{fig:LargeNmanycurves}}
\end{figure}

In Figs. \ref{fig:QMPQMmanycurves} and  \ref{fig:LargeNmanycurves} we show the scaling behavior of the order parameter for several values  of the symmetry breaking term $H/H_0=(m_\pi/m_{\pi ~{phys}})^2$. The leading-order corrections to  the scaling functions of the QM, PQM  and $O(N)$ models were discussed in the preceding section.

A comparison of the (mean-field) scaling properties of the order parameter in the QM and PQM models, shown in  Fig. \ref{fig:QMPQMmanycurves}, shows   that  the coupling of quarks to the   Polyakov loop  enhances the scaling violation.  This is particularly apparent in the broken phase,  where the Polyakov loop  expectation value differs appreciably from unity. Nevertheless, up to the physical pion mass, both models  are still in the scaling regime  of the underlying second-order phase transition, as also found  in LQCD.

In the  scaling plot of the QM model, shown in   Fig. \ref{fig:QMPQMmanycurves}, there are     two distinct points, where  the  curves for different values of the pion mass cross. This behavior was anticipated in our discussion of Landau theory in Sec.~\ref{sec:Landau-theory}. By substituting   the  coefficients of  the Landau thermodynamic potential obtained in the QM model, given in Eq. (\ref{eq:HTEcoeffs}), into Eqs. (\ref{eq:crossingalpha},\ref{eq:crossingxz}), we obtain the following  $(z,x)$ coordinates of the crossing points
\begin{equation}
	C_1=\{-1.31,1.42\},\quad\quad C_2=\{1.50,0.55\},
\end{equation}
in agreement with the numerical results shown in Fig.~\ref{fig:QMPQMmanycurves}. Obviously, the crossing points are independent of the pion mass only as long as the subleading scaling violating terms are negligible.

In the PQM model, shown in the right panel of Fig.~\ref{fig:QMPQMmanycurves}, the location of the crossing points depends on the strength of the symmetry breaking field for values of the pion mass below the physical one. This indicates that in the PQM model, the convergence of the expansion in powers of the symmetry breaking field  in Eq. (\ref{eq:Landau_mEOS}) is worse than in the QM model. The above  behavior  can be linked to the coupling of quarks with  the Polyakov loop.   In the low-temperature phase, the quark fluctuations are suppressed by the Polyakov loop, which results in a weaker dependence of the chiral condensate on the temperature. Consequently, close to the critical point, the chiral restoration as a function of temperature in the PQM model is sharper than in the QM model. This implies,  that the size of the scaling window is reduced, and that deviations from scaling are larger in the PQM than in the QM model.

The difference in strength of the scaling violation found in  the QM and PQM models is even more  pronounced in the $O(N)$ sigma model.
As shown in  Fig.~\ref{fig:LargeNmanycurves}, the $O(N)$ model exhibits stronger  deviations from the universal scaling curve than the QM and  PQM models for the corresponding  strength of the symmetry breaking field.
Indeed, the scaling of the order parameter in the $O(N)$ model is  preserved  only for a very weak external field and   the  deviations from the universal line are substantial for the physical value of the pion mass. The qualitative differences in the universal scaling curves and in the strength of the scaling violation indicate that fluctuations of the meson fields, not accounted for in the mean-field models, play an important r\^{o}le in the determination of the magnetic equation of state.

In spite of the fact that deviations from the universal scaling curve are large in the $O(N)$ sigma model, the  lines with different pion masses cross at a unique point. This suggests  that close to the critical temperature, the subleading corrections in the magnetic  equation of state are negligible up to the physical value of the pion mass.   Applying the procedure  discussed in the previous section, one  finds that    $(z,x)$ coordinates of this  crossing point  appear at $(-0.15,1.06)$, in agreement with the numerical results shown in  Fig. \ref{fig:LargeNmanycurves}.

The strong violation of scaling obtained in this model is consistent with previous studies within the FRG approach~\cite{Braun:2010vd}. However, in contrast to the FRG results of~\cite{Braun:2010vd}, we do not observe the approximate scaling of the order parameter for pion masses $\sim 100$ MeV to a  nonuniversal line for $z>-1$.

\begin{table*}[t]
\setlength{\tabcolsep}{6mm}
\centering
\begin{tabular}{|l||c|c|c|c|c|c|c|}
\hline
Model & $H_0$ & $\sigma_0$ & $t_0$ & $h_0$ & $z_0$
\\ \hline \hline
QM $m_{\sigma}=400\, \textrm{MeV}$ & $m_{\pi}^2 f_{\pi}$ & $f_{\pi}$ & 0.34 & 6.99 & 10.64
\\ \hline
QM $m_{\sigma}=800\, \textrm{MeV}$ & $m_{\pi}^2 f_{\pi}$ & $f_{\pi}$ & 0.30 & 13.57 & 19.10
\\ \hline
PQM $m_{\sigma}=400\, \textrm{MeV}$ & $m_{\pi}^2 f_{\pi}$ & $f_{\pi}$ & 0.073 & 5.26 & 41.50
\\ \hline
LS $m_{\sigma}=400\, \textrm{MeV}$ & $m_{\pi}^2 f_{\pi}$ & $f_{\pi}$ & 0.74 & 2.22 & 1.85
\\ \hline
LS $m_{\sigma}=800\, \textrm{MeV}$ & $m_{\pi}^2 f_{\pi}$ & $f_{\pi}$ & 0.57 & 1.69 & 2.18
\\ \hline
QM FRG \cite{Braun:2010vd} & 1 & 1 & 23.86 GeV/$T_c$ & 346.4 $\mathrm{GeV}^3$ & 2.69
\\ \hline
Lattice (p4) $N_{\tau}=4$ \cite{Kaczmarek:2011zz} & $\frac{m_{\pi}^2 f_{\pi} m_s}{m_l}$ & $\frac{T_c^4 f_{\pi}}{m_s \left\langle \overline{\psi}\psi \right\rangle_l^{T=0}}$ & 0.00407 & 0.00295 & 53.92
\\ \hline
Lattice (p4)  $N_{\tau}=8$ \cite{Kaczmarek:2011zz} & $\frac{m_{\pi}^2 f_{\pi} m_s}{m_l}$ & $\frac{T_c^4 f_{\pi}}{m_s \left\langle \overline{\psi}\psi \right\rangle_l^{T=0}}$ & 0.00271 & 0.00048 & 27.27
\\ \hline
\end{tabular}
\caption{Comparison of the nonuniversal constants $t_0$, $h_0$ and $z_0$ in different theories. The normalization of the order parameter $\sigma_0$ and the external field $H_0$  differs; thus, the $t_0$ and $h_0$ values cannot be directly compared between  different models. The $z_0$ column contains converted values to our normalization convention;  thus, the results of different models can be directly compared. \label{tab:t0h0z0}}
\end{table*}

\subsection{Scaling violation and  nonuniversal parameters \label{sec:comparison}}

In previous sections we studied the leading-order corrections to the magnetic equation of state and scaling functions in the mean-field approximation to the QM and PQM models. Clearly such a calculation cannot reproduce the universal properties of the $O(4)$ criticality expected in QCD (in three dimensions). However, this can be achieved by systematically including fluctuations of the meson fields e.g. within the FRG approach. In this context we note that the mean-field approximation does reproduce the universal properties of the $O(4)$ model in four dimensions. The difference between the mean-field approach and $O(4)$, or equivalently between $O(4)$ in three and four dimensions, is illustrated in Fig. 2 on the level of the scaling functions. The scaling function for
the $O(N\to\infty)$ sigma model, which belongs to another universality class in three dimensions, is also shown in Fig. 2. In spite of these differences, the mean-field models and the $(1/N)$ expansion of the $O(N)$ model allow us to explore the scaling violation in a transparent framework and to illustrate general features
of the magnetic equation of state, which are expected to be independent of the universality class.

The differences in the strength of the scaling violation seen in Figs. \ref{fig:QMPQMmanycurves} and \ref{fig:LargeNmanycurves}  are connected with  very different values of the
nonuniversal parameters $t_0$, $h_0$ and $z_0$. This is seen in Table \ref{tab:t0h0z0}, where we summarize their values in the present model calculations and in the previous studies of the magnetic equation of state within the FRG approach,   as well as in $(2+1)$-flavor  LQCD.
In the QM model and  $O(N=\infty)$ linear sigma model,   explicit expressions for $h_0$ and $t_0$ are given in Eqs. (\ref{eq:MF_t0h0}) and  (\ref{eq:Oinf_t0h0}). In the PQM model these constants were obtained numerically, by fitting the order parameter to the asymptotic scaling laws Eq. (\ref{eq:h0t0def}).

Clearly,  the  values of these nonuniversal parameters are not only model dependent, but are also influenced  by   the normalization convention of the external field and the  order parameter. The constant $z_0$, however,  does not depend  on the choice of the normalization of the order parameter.  The values of  $z_0$ given in Table \ref{tab:t0h0z0}  can be directly compared between different models, since  they were recomputed  with  the same normalization of the external field.

\begin{figure*}[tb]
\centering
   \includegraphics[width=0.48\linewidth]{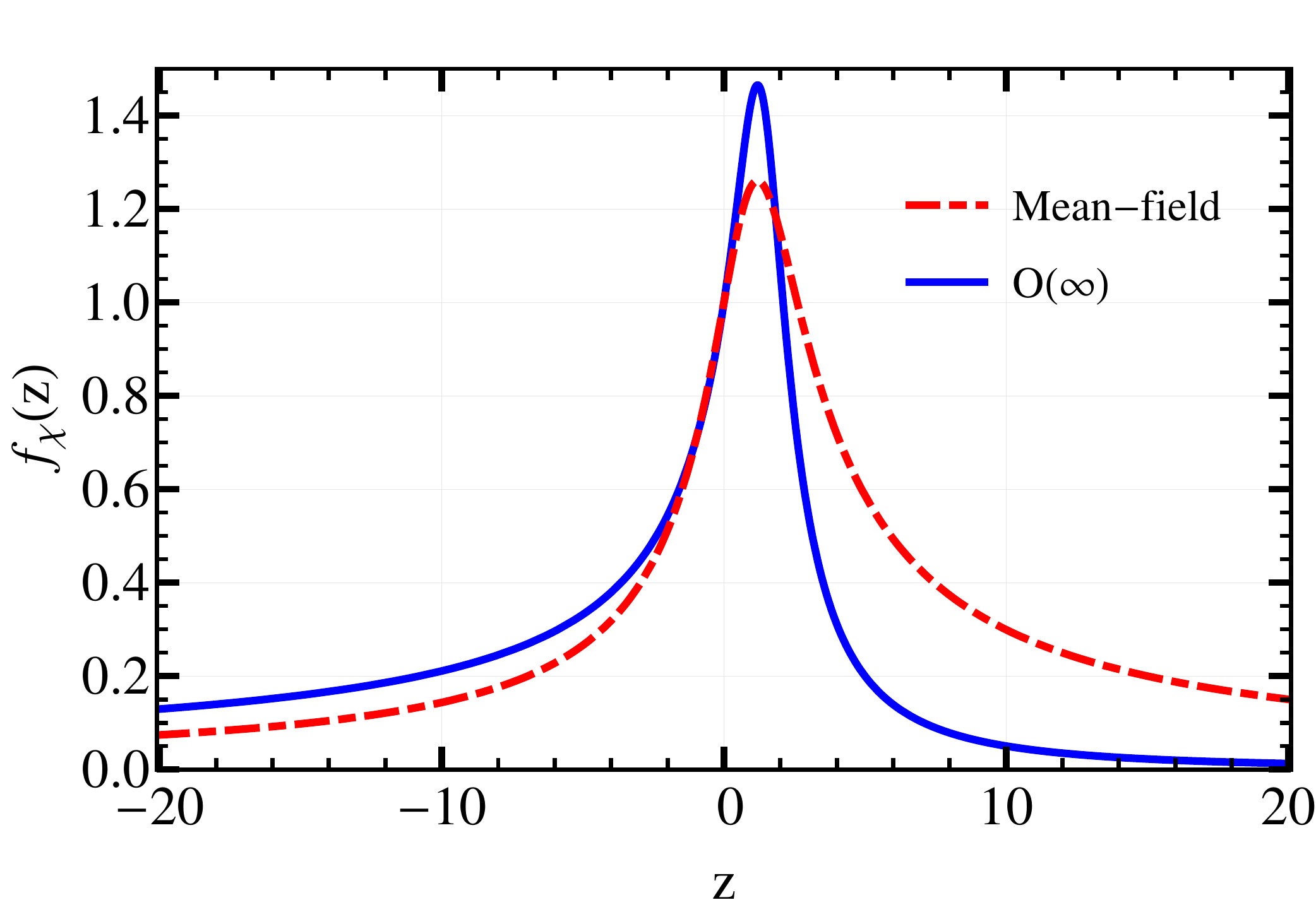}
   \includegraphics[width=0.48\linewidth]{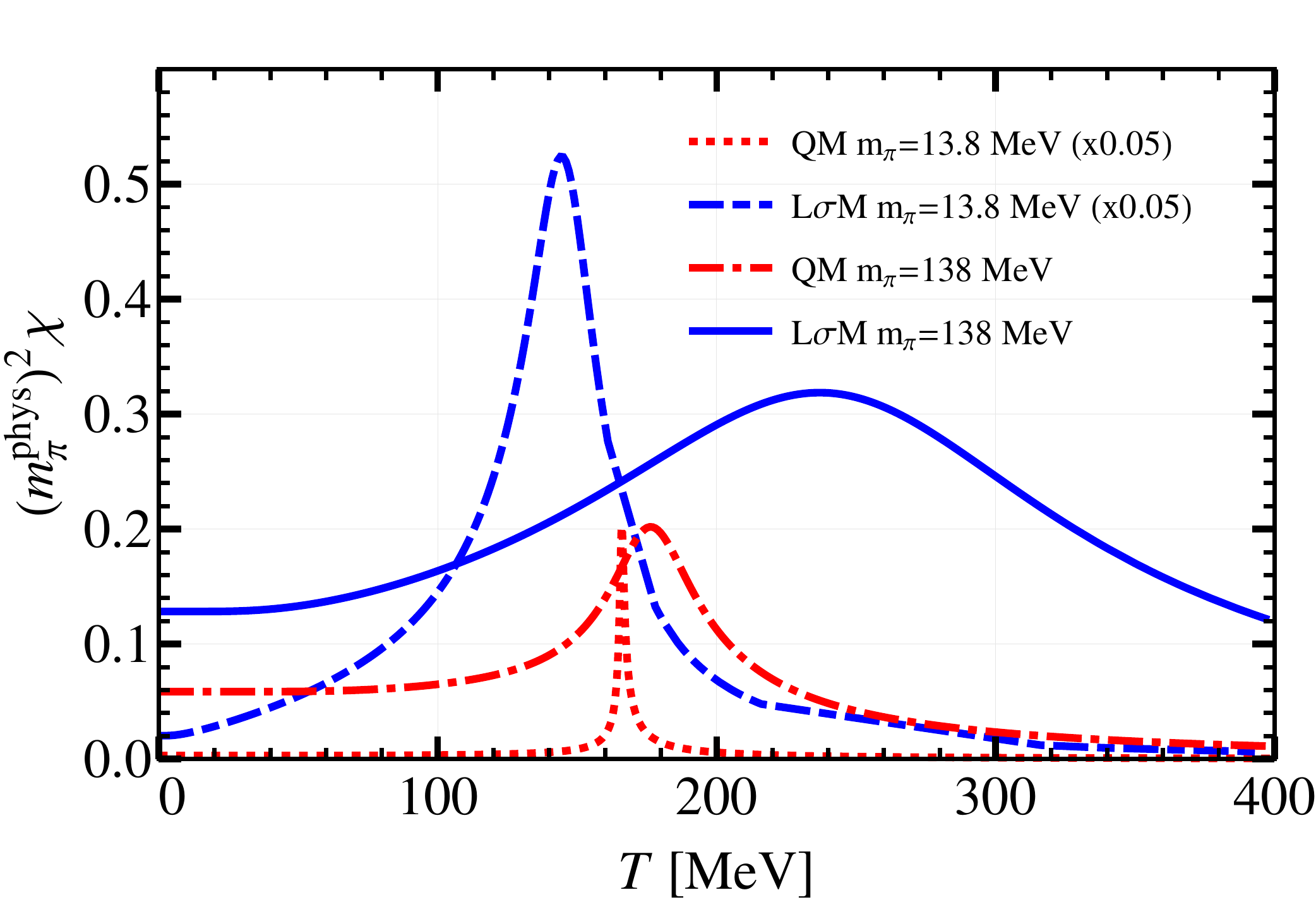}
   \caption{Left:  The universal scaling part  of the chiral susceptibility in the QM model calculated within  the mean-field dynamics and in  the $O(N)$ LS model obtained in the $N\to\infty$ limit within the high-temperature expansion. Right:  The chiral  susceptibilities in the LS model in the $N\to \infty$ limit and in the QM model under the mean-field approximation calculated  at  physical and at ten times lower pion mass. \label{fig:susc}}
\end{figure*}

\begin{figure}[b]
	\centering
	\includegraphics[width=0.98\linewidth]{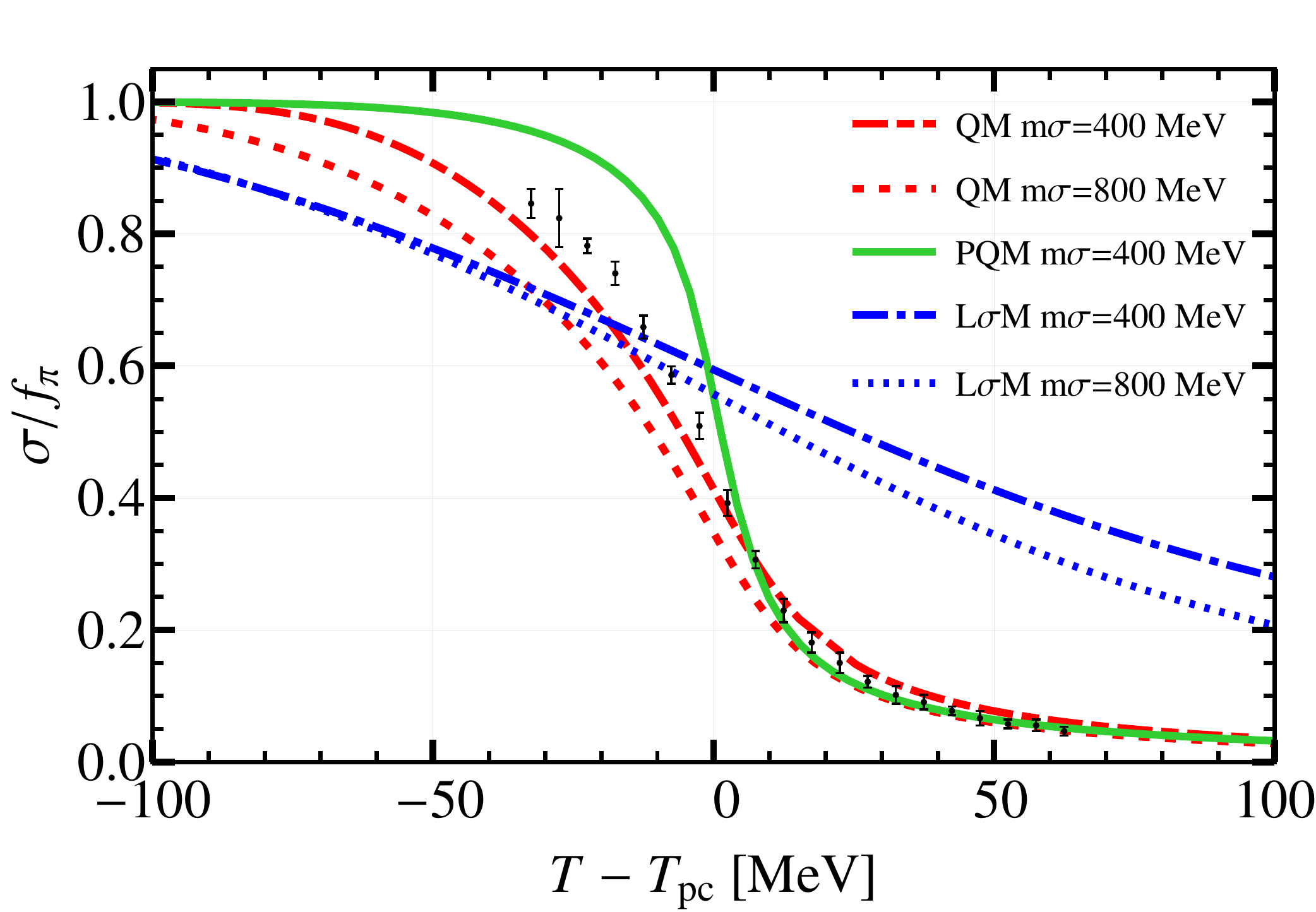}
	\caption{The chiral order parameters in the QM and LS  models  calculated with different inputs for the vacuum sigma mass.  The  data points are lattice results from Ref. \cite{Borsanyi:2010bp}. The pseudocritical temperature  $T_{pc}$ for these data is taken as the inflection point of the order parameter. \label{fig:melting}}
\end{figure}

From Table~\ref{tab:t0h0z0}  it is clear that the $z_0$ values obtained in the mean-field models  are  roughly compatible with the lattice results. On the other hand, the effective models with bosonic fluctuations yield a much smaller $z_0$. We note that here we are comparing 2-flavor model calculations with (2+1)-flavor lattice QCD simulations. Such a comparison makes sense, since the strange quark remains massive at the chiral transition and hence contributes only to the regular part of the free energy. This leads to small reduction of the chiral transition temperature, but has only a minor effect on the critical properties. Thus, for the purposes of this exploratory study, 
the neglect of the strange quark is reasonable.

As discussed in Sec.~\ref*{sec:Universality}, the nonuniversal parameter $z_0$ influences the critical properties of relevant observables in the crossover regime.
In particular, as shown in Eqs. (\ref{eq:Tpcdependence}) and (\ref{eq:crossoverwidth}),  the constant $z_0$ determines  the width of the transition region and the   peak position of the  order parameter susceptibility $\chi$. In Fig. \ref{fig:susc} we show the chiral susceptibility,  computed in the QM and in $O(N\to\infty)$ linear sigma models. The left panel shows the universal part of the chiral susceptibility, whereas the right one depicts the  temperature dependence of $\chi$ for $m_\pi/m_{\pi\,{\rm phys}}=0.1$ and $1$. Although the scaling functions of these models correspond to   different universality classes, they are quantitatively  rather similar. This, however  is not the case for $\chi(T,m_\pi)$, since owing to the difference in the values of $z_0$, the
crossover region  in the LS model is considerably  wider and the shift in the pseudocritical temperature with increasing pion mass is larger than  in the QM model.

Moreover, due to comparable values of $z_0$ in the mean-field models and in LQCD, the melting of the chiral condensate in LQCD should be better described by the QM and PQM model than by the LS model. This is indeed seen in Fig. \ref{fig:melting}, where we compare the LQCD data with model results.
Models where bosonic fluctuations are included~\cite{Braun:2010vd}, such as the LS model in the $N\to\infty$ limit,  yield a much smaller  value of  $z_0$ than  LQCD calculations. Hence,  the reduction of the chiral condensate at the crossover transition is much smoother than in LQCD. Such a broadening of the transition region, when meson fluctuations are included, was observed also in the QM and PQM model mean-field and FRG calculations of Ref.~\cite{Skokov:2010wb}.

By comparing different model results obtained in the present studies, together with previous FRG findings in the PQM model and LQCD results,  we could confirm the r\^{o}le of the parameter $z_0$,  which, using Eq. (\ref{eq:crossoverwidth}), determines the width of the transition in physical units. Again, this correlation is independent of the universality class, modulo minor variations in $z_p$. Thus, theories with a large $z_0$, roughly comparable with the lattice results, exhibit a relatively narrow transition region, as found in LQCD. This is the case for the QM and, in particular, for the PQM models in the mean-field approximation. On the other hand,  theories with a small value of $z_0$, like the QM model with mesonic fluctuations included as well as the $O(N)$ sigma model at large $N$,  exhibit a much smoother transition. Consequently, a viable effective model for the critical chiral dynamics of QCD should exhibit a value for $z_0$ comparable to that obtained in LQCD.
	
Moreover, we find that the scaling window depends on the parameter $h_0$ and on the nonsingular background, which in the Landau model, to leading order, is determined by the sixth-order coupling $c_0$. Thus, we conclude that in a given model the scaling window can be tuned to agree with lattice QCD by varying the nonuniversal parameter $h_0$ and the strength of the effective sixth-order coupling.
	
We note that adjusting model parameters to lattice QCD results for certain nonuniversal quantities does not guarantee that other nonuniversal quantities are reproduced by the model. However, for modeling the effect of critical fluctuations at the chiral transition, the width of the transition region and the size of the scaling window are, besides the universality class, the most important criteria for discriminating between models. Our study indicates how effective models can be tuned so that these key quantities are reproduced.

\section{Conclusion \label{sec:Conclusion}}

We have discussed universal and nonuniversal aspects of the chiral phase transition and the corresponding  magnetic equation of state  in different effective models of QCD. The critical properties of the QM and PQM models were explored in the mean-field approximation, where only fermionic fluctuations are accounted for, and compared to those  of  a purely bosonic theory, the $O(N)$ linear sigma model in the $N\to\infty$ limit. In the QM  and LS models the magnetic equation of state was computed analytically within  the high-temperature expansion.  The effects of a gluonic  background on the nonuniversal scaling parameters  were assessed within the  PQM model.

We have analyzed the scaling violation at nonzero quark masses  in the context of recent LQCD results, which indicate that, at a physical pion mass, QCD lies in the scaling regime  of the underlying second-order phase transition.
We showed that to understand the chiral critical properties of QCD it is not enough to have a model in the same universality class, but the model in question should approach criticality in a similar manner. We quantified this with dimensionless, nonuniversal parameters $t_0$, $h_0$ and $z_0$ that connect the physical quark mass and temperature scales with the dimensionless scaling variables of the universality class. We found that these nonuniversal quantities differ significantly from model to model and this influences the size of the order parameter scaling window.

In the QM and PQM models, the scaling violating contributions to the order parameter  were found to remain small up to the physical pion mass.  This is in qualitative agreement with the scaling behavior found in LQCD at the chiral crossover transition.

On the other hand, in the $O(N)$ LS model, which we solved in the $N\to\infty$ limit,
we found that the fluctuations of the meson fields  yield a much stronger scaling violation than that obtained in the QM and PQM models, which in the mean-field approximation accounts only for fluctuations of fermions. In particular, we observed that the order parameter in the LS model follows the universal scaling law only for
very small values of the pion mass.  Consequently, at physical pion mass,
the chiral condensate in the LS model  exhibits  substantial violation of the universal scaling law.

The very different scaling behavior of  these models was linked  to  very different values of the nonuniversal  scaling parameter  $z_0$. In the QM and PQM model,   $z_0$ was  found to be roughly compatible with that obtained in LQCD, whereas in  the LS model this parameter is almost   an order of magnitude smaller. The value of  $z_0$ is also reflected in the width  of the crossover transition and the shift in the peak position of the chiral susceptibility with increasing pion mass.
This analysis  indicates  that  models where bosonic fluctuations are accounted for,  tend to have a small $z_0$, a broad peak in the chiral susceptibility and a narrow   critical region.

From general considerations in Landau theory we have obtained a connection between distinctive features of the scaling violation and specific
properties of the coefficients of the effective potential. This provides a
framework for discussing general characteristics of the scaling violation in
terms of the model parameters and in particular to understand how the
scaling function approaches the universal scaling curve. We found that, depending on the temperature dependence of the coefficients of the effective
potential, the magnetic equation of state may exhibit a nontrivial structure with common  crossing points for different values of the symmetry breaking  field $H$.  We have quantified these properties in the QM, PQM and LS models.
In the QM model, there are two distinct points on the universal scaling line, where to leading order in $H$,  all curves cross. In the LS model, there is only one such point, while in the PQM model the crossing of the universal scaling line is $H$ dependent already for pion masses well below its physical value. We presented a straightforward interpretation of these features, based on general considerations  derived within Landau theory.

Finally, we argued that the critical properties of QCD, namely the width of the chiral transition and the size of the scaling window, can be reproduced by tuning the nonuniversal parameters $z_0$ and $h_0$ and the strength of the effective sixth-order coupling. Our calculations indicate that this is indeed the case in the mean-field models and in the large-$N$ linear sigma model considered.\newline

\section*{Acknowledgments}

The work of B.F. and K.R.  was partly supported
by the Extreme Matter Institute EMMI.
K. R. also  acknowledges  partial  supports of the Polish Science Center  (NCN) under Maestro Grant No. DEC-2013/10/A/ST2/00106, and    the  U.S.
Department  of  Energy  under  Grant  No.  DE-FG02-05ER41367.
W.T. is grateful to GSI for the hospitality during the Summer Student Programme.
G. A. acknowledges the support of  the Hessian LOEWE initiative
through the Helmholtz International Center for FAIR (HIC for FAIR).

\bibliography{refs}

\begin{thebibliography}{60}%
\makeatletter
\providecommand \@ifxundefined [1]{%
 \@ifx{#1\undefined}
}%
\providecommand \@ifnum [1]{%
 \ifnum #1\expandafter \@firstoftwo
 \else \expandafter \@secondoftwo
 \fi
}%
\providecommand \@ifx [1]{%
 \ifx #1\expandafter \@firstoftwo
 \else \expandafter \@secondoftwo
 \fi
}%
\providecommand \natexlab [1]{#1}%
\providecommand \enquote  [1]{``#1''}%
\providecommand \bibnamefont  [1]{#1}%
\providecommand \bibfnamefont [1]{#1}%
\providecommand \citenamefont [1]{#1}%
\providecommand \href@noop [0]{\@secondoftwo}%
\providecommand \href [0]{\begingroup \@sanitize@url \@href}%
\providecommand \@href[1]{\@@startlink{#1}\@@href}%
\providecommand \@@href[1]{\endgroup#1\@@endlink}%
\providecommand \@sanitize@url [0]{\catcode `\\12\catcode `\$12\catcode
  `\&12\catcode `\#12\catcode `\^12\catcode `\_12\catcode `\%12\relax}%
\providecommand \@@startlink[1]{}%
\providecommand \@@endlink[0]{}%
\providecommand \url  [0]{\begingroup\@sanitize@url \@url }%
\providecommand \@url [1]{\endgroup\@href {#1}{\urlprefix }}%
\providecommand \urlprefix  [0]{URL }%
\providecommand \Eprint [0]{\href }%
\providecommand \doibase [0]{http://dx.doi.org/}%
\providecommand \selectlanguage [0]{\@gobble}%
\providecommand \bibinfo  [0]{\@secondoftwo}%
\providecommand \bibfield  [0]{\@secondoftwo}%
\providecommand \translation [1]{[#1]}%
\providecommand \BibitemOpen [0]{}%
\providecommand \bibitemStop [0]{}%
\providecommand \bibitemNoStop [0]{.\EOS\space}%
\providecommand \EOS [0]{\spacefactor3000\relax}%
\providecommand \BibitemShut  [1]{\csname bibitem#1\endcsname}%
\let\auto@bib@innerbib\@empty
\bibitem [{\citenamefont {Friman}\ \emph {et~al.}(2011)\citenamefont {Friman},
  \citenamefont {Hohne}, \citenamefont {Knoll}, \citenamefont {Leupold},
  \citenamefont {Randrup}, \citenamefont {Rapp},\ and\ \citenamefont
  {Senger}}]{Friman:2011zz}%
  \BibitemOpen
  \bibfield  {author} {\bibinfo {author} {\bibfnamefont {B.}~\bibnamefont
  {Friman}}, \bibinfo {author} {\bibfnamefont {C.}~\bibnamefont {Hohne}},
  \bibinfo {author} {\bibfnamefont {J.}~\bibnamefont {Knoll}}, \bibinfo
  {author} {\bibfnamefont {S.}~\bibnamefont {Leupold}}, \bibinfo {author}
  {\bibfnamefont {J.}~\bibnamefont {Randrup}}, \bibinfo {author} {\bibfnamefont
  {R.}~\bibnamefont {Rapp}}, \ and\ \bibinfo {author} {\bibfnamefont
  {P.}~\bibnamefont {Senger}},\ }\href {\doibase 10.1007/978-3-642-13293-3}
  {\bibfield  {journal} {\bibinfo  {journal} {Lect. Notes Phys.}\ }\textbf
  {\bibinfo {volume} {814}} (\bibinfo {year} {2011}),\
  10.1007/978-3-642-13293-3}\BibitemShut {NoStop}%
\bibitem [{\citenamefont {Fukushima}\ and\ \citenamefont
  {Hatsuda}(2011)}]{Fukushima:2010bq}%
  \BibitemOpen
  \bibfield  {author} {\bibinfo {author} {\bibfnamefont {K.}~\bibnamefont
  {Fukushima}}\ and\ \bibinfo {author} {\bibfnamefont {T.}~\bibnamefont
  {Hatsuda}},\ }\href {\doibase 10.1088/0034-4885/74/1/014001} {\bibfield
  {journal} {\bibinfo  {journal} {Rept. Prog. Phys.}\ }\textbf {\bibinfo
  {volume} {74}},\ \bibinfo {pages} {014001} (\bibinfo {year}
  {2011})}\BibitemShut {NoStop}%
\bibitem [{\citenamefont {Fukushima}\ and\ \citenamefont
  {Sasaki}(2013)}]{Fukushima:2013rx}%
  \BibitemOpen
  \bibfield  {author} {\bibinfo {author} {\bibfnamefont {K.}~\bibnamefont
  {Fukushima}}\ and\ \bibinfo {author} {\bibfnamefont {C.}~\bibnamefont
  {Sasaki}},\ }\href {\doibase 10.1016/j.ppnp.2013.05.003} {\bibfield
  {journal} {\bibinfo  {journal} {Prog. Part. Nucl. Phys.}\ }\textbf {\bibinfo
  {volume} {72}},\ \bibinfo {pages} {99} (\bibinfo {year} {2013})}\BibitemShut
  {NoStop}%
\bibitem [{\citenamefont {Pisarski}\ and\ \citenamefont
  {Wilczek}(1984)}]{Pisarski:1983ms}%
  \BibitemOpen
  \bibfield  {author} {\bibinfo {author} {\bibfnamefont {R.~D.}\ \bibnamefont
  {Pisarski}}\ and\ \bibinfo {author} {\bibfnamefont {F.}~\bibnamefont
  {Wilczek}},\ }\href {\doibase 10.1103/PhysRevD.29.338} {\bibfield  {journal}
  {\bibinfo  {journal} {Phys. Rev. D}\ }\textbf {\bibinfo {volume} {29}},\
  \bibinfo {pages} {338} (\bibinfo {year} {1984})}\BibitemShut {NoStop}%
\bibitem [{\citenamefont {Aoki}\ \emph {et~al.}(2006)\citenamefont {Aoki},
  \citenamefont {Endrodi}, \citenamefont {Fodor}, \citenamefont {Katz},\ and\
  \citenamefont {Szabo}}]{Aoki:2006we}%
  \BibitemOpen
  \bibfield  {author} {\bibinfo {author} {\bibfnamefont {Y.}~\bibnamefont
  {Aoki}}, \bibinfo {author} {\bibfnamefont {G.}~\bibnamefont {Endrodi}},
  \bibinfo {author} {\bibfnamefont {Z.}~\bibnamefont {Fodor}}, \bibinfo
  {author} {\bibfnamefont {S.~D.}\ \bibnamefont {Katz}}, \ and\ \bibinfo
  {author} {\bibfnamefont {K.~K.}\ \bibnamefont {Szabo}},\ }\href {\doibase
  10.1038/nature05120} {\bibfield  {journal} {\bibinfo  {journal} {Nature}\
  }\textbf {\bibinfo {volume} {443}},\ \bibinfo {pages} {675} (\bibinfo {year}
  {2006})}\BibitemShut {NoStop}%
\bibitem [{\citenamefont {Ejiri}\ \emph {et~al.}(2009)\citenamefont {Ejiri},
  \citenamefont {Karsch}, \citenamefont {Laermann}, \citenamefont {Miao},
  \citenamefont {Mukherjee} \emph {et~al.}}]{Ejiri:2009ac}%
  \BibitemOpen
  \bibfield  {author} {\bibinfo {author} {\bibfnamefont {S.}~\bibnamefont
  {Ejiri}}, \bibinfo {author} {\bibfnamefont {F.}~\bibnamefont {Karsch}},
  \bibinfo {author} {\bibfnamefont {E.}~\bibnamefont {Laermann}}, \bibinfo
  {author} {\bibfnamefont {C.}~\bibnamefont {Miao}}, \bibinfo {author}
  {\bibfnamefont {S.}~\bibnamefont {Mukherjee}},  \emph {et~al.},\ }\href
  {\doibase 10.1103/PhysRevD.80.094505} {\bibfield  {journal} {\bibinfo
  {journal} {Phys. Rev. D}\ }\textbf {\bibinfo {volume} {80}},\ \bibinfo
  {pages} {094505} (\bibinfo {year} {2009})}\BibitemShut {NoStop}%
\bibitem [{\citenamefont {Kaczmarek}\ \emph {et~al.}(2011)\citenamefont
  {Kaczmarek}, \citenamefont {Karsch}, \citenamefont {Laermann}, \citenamefont
  {Miao}, \citenamefont {Mukherjee} \emph {et~al.}}]{Kaczmarek:2011zz}%
  \BibitemOpen
  \bibfield  {author} {\bibinfo {author} {\bibfnamefont {O.}~\bibnamefont
  {Kaczmarek}}, \bibinfo {author} {\bibfnamefont {F.}~\bibnamefont {Karsch}},
  \bibinfo {author} {\bibfnamefont {E.}~\bibnamefont {Laermann}}, \bibinfo
  {author} {\bibfnamefont {C.}~\bibnamefont {Miao}}, \bibinfo {author}
  {\bibfnamefont {S.}~\bibnamefont {Mukherjee}},  \emph {et~al.},\ }\href
  {\doibase 10.1103/PhysRevD.83.014504} {\bibfield  {journal} {\bibinfo
  {journal} {Phys. Rev. D}\ }\textbf {\bibinfo {volume} {83}},\ \bibinfo
  {pages} {014504} (\bibinfo {year} {2011})}\BibitemShut {NoStop}%
\bibitem [{\citenamefont {Ding}\ \emph {et~al.}(2014)\citenamefont {Ding},
  \citenamefont {Bazavov}, \citenamefont {Karsch}, \citenamefont {Maezawa},
  \citenamefont {Mukherjee} \emph {et~al.}}]{Ding:2013lfa}%
  \BibitemOpen
  \bibfield  {author} {\bibinfo {author} {\bibfnamefont {H.-T.}\ \bibnamefont
  {Ding}}, \bibinfo {author} {\bibfnamefont {A.}~\bibnamefont {Bazavov}},
  \bibinfo {author} {\bibfnamefont {F.}~\bibnamefont {Karsch}}, \bibinfo
  {author} {\bibfnamefont {Y.}~\bibnamefont {Maezawa}}, \bibinfo {author}
  {\bibfnamefont {S.}~\bibnamefont {Mukherjee}},  \emph {et~al.},\ }\href
  {http://inspirehep.net/record/1266965/files/arXiv:1312.0119.pdf} {\bibfield
  {journal} {\bibinfo  {journal} {Proc. Sci. LATTICE2013}\ ,\ \bibinfo {pages}
  {157}} (\bibinfo {year} {2014})}\BibitemShut {NoStop}%
\bibitem [{\citenamefont {Ding}\ and\ \citenamefont
  {Hegde}(2016)}]{Ding:2015pmg}%
  \BibitemOpen
  \bibfield  {author} {\bibinfo {author} {\bibfnamefont {H.-T.}\ \bibnamefont
  {Ding}}\ and\ \bibinfo {author} {\bibfnamefont {P.}~\bibnamefont {Hegde}}
  (\bibinfo {collaboration} {Bielefeld-BNL-CCNU}),\ }\href
  {http://inspirehep.net/record/1402370/files/arXiv:1511.00553.pdf} {\bibfield
  {journal} {\bibinfo  {journal} {Proc. Sci. LATTICE2015}\ ,\ \bibinfo {pages}
  {161}} (\bibinfo {year} {2016})}\BibitemShut {NoStop}%
\bibitem [{\citenamefont {Zinn-Justin}(2002)}]{zinn2002quantum}%
  \BibitemOpen
  \bibfield  {author} {\bibinfo {author} {\bibfnamefont {J.}~\bibnamefont
  {Zinn-Justin}},\ }\href@noop {} {\emph {\bibinfo {title} {Quantum Field
  Theory and Critical Phenomena}}},\ International series of monographs on
  physics\ (\bibinfo  {publisher} {Clarendon Press},\ \bibinfo {year} {Oxford,
  2002})\BibitemShut {NoStop}%
\bibitem [{\citenamefont {D'Elia}\ \emph {et~al.}(2005)\citenamefont {D'Elia},
  \citenamefont {Di~Giacomo},\ and\ \citenamefont {Pica}}]{D'Elia:2005bv}%
  \BibitemOpen
  \bibfield  {author} {\bibinfo {author} {\bibfnamefont {M.}~\bibnamefont
  {D'Elia}}, \bibinfo {author} {\bibfnamefont {A.}~\bibnamefont {Di~Giacomo}},
  \ and\ \bibinfo {author} {\bibfnamefont {C.}~\bibnamefont {Pica}},\ }\href
  {\doibase 10.1103/PhysRevD.72.114510} {\bibfield  {journal} {\bibinfo
  {journal} {Phys. Rev. D}\ }\textbf {\bibinfo {volume} {72}},\ \bibinfo
  {pages} {114510} (\bibinfo {year} {2005})}\BibitemShut {NoStop}%
\bibitem [{\citenamefont {Bonati}\ \emph {et~al.}(2009)\citenamefont {Bonati},
  \citenamefont {Cossu}, \citenamefont {D'Elia}, \citenamefont {Di~Giacomo},\
  and\ \citenamefont {Pica}}]{Bonati:2009zz}%
  \BibitemOpen
  \bibfield  {author} {\bibinfo {author} {\bibfnamefont {C.}~\bibnamefont
  {Bonati}}, \bibinfo {author} {\bibfnamefont {G.}~\bibnamefont {Cossu}},
  \bibinfo {author} {\bibfnamefont {M.}~\bibnamefont {D'Elia}}, \bibinfo
  {author} {\bibfnamefont {A.}~\bibnamefont {Di~Giacomo}}, \ and\ \bibinfo
  {author} {\bibfnamefont {C.}~\bibnamefont {Pica}},\ }\href {\doibase
  10.1016/j.nuclphysa.2009.01.060} {\bibfield  {journal} {\bibinfo  {journal}
  {Nucl. Phys.}\ }\textbf {\bibinfo {volume} {A820}},\ \bibinfo {pages} {243C}
  (\bibinfo {year} {2009})}\BibitemShut {NoStop}%
\bibitem [{\citenamefont {Bonati}\ \emph {et~al.}(2014)\citenamefont {Bonati},
  \citenamefont {de~Forcrand}, \citenamefont {D'Elia}, \citenamefont
  {Philipsen},\ and\ \citenamefont {Sanfilippo}}]{Bonati:2014kpa}%
  \BibitemOpen
  \bibfield  {author} {\bibinfo {author} {\bibfnamefont {C.}~\bibnamefont
  {Bonati}}, \bibinfo {author} {\bibfnamefont {P.}~\bibnamefont {de~Forcrand}},
  \bibinfo {author} {\bibfnamefont {M.}~\bibnamefont {D'Elia}}, \bibinfo
  {author} {\bibfnamefont {O.}~\bibnamefont {Philipsen}}, \ and\ \bibinfo
  {author} {\bibfnamefont {F.}~\bibnamefont {Sanfilippo}},\ }\href {\doibase
  10.1103/PhysRevD.90.074030} {\bibfield  {journal} {\bibinfo  {journal} {Phys.
  Rev. D}\ }\textbf {\bibinfo {volume} {90}},\ \bibinfo {pages} {074030}
  (\bibinfo {year} {2014})}\BibitemShut {NoStop}%
\bibitem [{\citenamefont {Gell-Mann}\ and\ \citenamefont
  {Levy}(1960)}]{GellMann:1960np}%
  \BibitemOpen
  \bibfield  {author} {\bibinfo {author} {\bibfnamefont {M.}~\bibnamefont
  {Gell-Mann}}\ and\ \bibinfo {author} {\bibfnamefont {M.}~\bibnamefont
  {Levy}},\ }\href {\doibase 10.1007/BF02859738} {\bibfield  {journal}
  {\bibinfo  {journal} {Nuovo Cim.}\ }\textbf {\bibinfo {volume} {16}},\
  \bibinfo {pages} {705} (\bibinfo {year} {1960})}\BibitemShut {NoStop}%
\bibitem [{\citenamefont {Schaefer}\ \emph {et~al.}(2007)\citenamefont
  {Schaefer}, \citenamefont {Pawlowski},\ and\ \citenamefont
  {Wambach}}]{Schaefer:2007pw}%
  \BibitemOpen
  \bibfield  {author} {\bibinfo {author} {\bibfnamefont {B.-J.}\ \bibnamefont
  {Schaefer}}, \bibinfo {author} {\bibfnamefont {J.~M.}\ \bibnamefont
  {Pawlowski}}, \ and\ \bibinfo {author} {\bibfnamefont {J.}~\bibnamefont
  {Wambach}},\ }\href {\doibase 10.1103/PhysRevD.76.074023} {\bibfield
  {journal} {\bibinfo  {journal} {Phys. Rev. D}\ }\textbf {\bibinfo {volume}
  {76}},\ \bibinfo {pages} {074023} (\bibinfo {year} {2007})}\BibitemShut
  {NoStop}%
\bibitem [{\citenamefont {Schaefer}\ \emph {et~al.}(2010)\citenamefont
  {Schaefer}, \citenamefont {Wagner},\ and\ \citenamefont
  {Wambach}}]{Schaefer:2009ui}%
  \BibitemOpen
  \bibfield  {author} {\bibinfo {author} {\bibfnamefont {B.-J.}\ \bibnamefont
  {Schaefer}}, \bibinfo {author} {\bibfnamefont {M.}~\bibnamefont {Wagner}}, \
  and\ \bibinfo {author} {\bibfnamefont {J.}~\bibnamefont {Wambach}},\ }\href
  {\doibase 10.1103/PhysRevD.81.074013} {\bibfield  {journal} {\bibinfo
  {journal} {Phys. Rev. D}\ }\textbf {\bibinfo {volume} {81}},\ \bibinfo
  {pages} {074013} (\bibinfo {year} {2010})}\BibitemShut {NoStop}%
\bibitem [{\citenamefont {Mizher}\ \emph {et~al.}(2010)\citenamefont {Mizher},
  \citenamefont {Chernodub},\ and\ \citenamefont {Fraga}}]{Mizher:2010zb}%
  \BibitemOpen
  \bibfield  {author} {\bibinfo {author} {\bibfnamefont {A.~J.}\ \bibnamefont
  {Mizher}}, \bibinfo {author} {\bibfnamefont {M.~N.}\ \bibnamefont
  {Chernodub}}, \ and\ \bibinfo {author} {\bibfnamefont {E.~S.}\ \bibnamefont
  {Fraga}},\ }\href {\doibase 10.1103/PhysRevD.82.105016} {\bibfield  {journal}
  {\bibinfo  {journal} {Phys. Rev. D}\ }\textbf {\bibinfo {volume} {82}},\
  \bibinfo {pages} {105016} (\bibinfo {year} {2010})}\BibitemShut {NoStop}%
\bibitem [{\citenamefont {Herbst}\ \emph {et~al.}(2011)\citenamefont {Herbst},
  \citenamefont {Pawlowski},\ and\ \citenamefont {Schaefer}}]{Herbst:2010rf}%
  \BibitemOpen
  \bibfield  {author} {\bibinfo {author} {\bibfnamefont {T.~K.}\ \bibnamefont
  {Herbst}}, \bibinfo {author} {\bibfnamefont {J.~M.}\ \bibnamefont
  {Pawlowski}}, \ and\ \bibinfo {author} {\bibfnamefont {B.-J.}\ \bibnamefont
  {Schaefer}},\ }\href {\doibase 10.1016/j.physletb.2010.12.003} {\bibfield
  {journal} {\bibinfo  {journal} {Phys. Lett. B}\ }\textbf {\bibinfo {volume}
  {696}},\ \bibinfo {pages} {58} (\bibinfo {year} {2011})}\BibitemShut
  {NoStop}%
\bibitem [{\citenamefont {Skokov}\ \emph
  {et~al.}(2010{\natexlab{a}})\citenamefont {Skokov}, \citenamefont {Stokic},
  \citenamefont {Friman},\ and\ \citenamefont {Redlich}}]{Skokov:2010wb}%
  \BibitemOpen
  \bibfield  {author} {\bibinfo {author} {\bibfnamefont {V.}~\bibnamefont
  {Skokov}}, \bibinfo {author} {\bibfnamefont {B.}~\bibnamefont {Stokic}},
  \bibinfo {author} {\bibfnamefont {B.}~\bibnamefont {Friman}}, \ and\ \bibinfo
  {author} {\bibfnamefont {K.}~\bibnamefont {Redlich}},\ }\href {\doibase
  10.1103/PhysRevC.82.015206} {\bibfield  {journal} {\bibinfo  {journal} {Phys.
  Rev. C}\ }\textbf {\bibinfo {volume} {82}},\ \bibinfo {pages} {015206}
  (\bibinfo {year} {2010}{\natexlab{a}})}\BibitemShut {NoStop}%
\bibitem [{\citenamefont {Skokov}\ \emph
  {et~al.}(2010{\natexlab{b}})\citenamefont {Skokov}, \citenamefont {Friman},
  \citenamefont {Nakano}, \citenamefont {Redlich},\ and\ \citenamefont
  {Schaefer}}]{Skokov:2010sf}%
  \BibitemOpen
  \bibfield  {author} {\bibinfo {author} {\bibfnamefont {V.}~\bibnamefont
  {Skokov}}, \bibinfo {author} {\bibfnamefont {B.}~\bibnamefont {Friman}},
  \bibinfo {author} {\bibfnamefont {E.}~\bibnamefont {Nakano}}, \bibinfo
  {author} {\bibfnamefont {K.}~\bibnamefont {Redlich}}, \ and\ \bibinfo
  {author} {\bibfnamefont {B.~J.}\ \bibnamefont {Schaefer}},\ }\href {\doibase
  10.1103/PhysRevD.82.034029} {\bibfield  {journal} {\bibinfo  {journal} {Phys.
  Rev. D}\ }\textbf {\bibinfo {volume} {82}},\ \bibinfo {pages} {034029}
  (\bibinfo {year} {2010}{\natexlab{b}})}\BibitemShut {NoStop}%
\bibitem [{\citenamefont {Skokov}\ \emph {et~al.}(2011)\citenamefont {Skokov},
  \citenamefont {Friman},\ and\ \citenamefont {Redlich}}]{Skokov:2010uh}%
  \BibitemOpen
  \bibfield  {author} {\bibinfo {author} {\bibfnamefont {V.}~\bibnamefont
  {Skokov}}, \bibinfo {author} {\bibfnamefont {B.}~\bibnamefont {Friman}}, \
  and\ \bibinfo {author} {\bibfnamefont {K.}~\bibnamefont {Redlich}},\ }\href
  {\doibase 10.1103/PhysRevC.83.054904} {\bibfield  {journal} {\bibinfo
  {journal} {Phys. Rev. C}\ }\textbf {\bibinfo {volume} {83}},\ \bibinfo
  {pages} {054904} (\bibinfo {year} {2011})}\BibitemShut {NoStop}%
\bibitem [{\citenamefont {Mitter}\ and\ \citenamefont
  {Schaefer}(2014)}]{Mitter:2013fxa}%
  \BibitemOpen
  \bibfield  {author} {\bibinfo {author} {\bibfnamefont {M.}~\bibnamefont
  {Mitter}}\ and\ \bibinfo {author} {\bibfnamefont {B.-J.}\ \bibnamefont
  {Schaefer}},\ }\href {\doibase 10.1103/PhysRevD.89.054027} {\bibfield
  {journal} {\bibinfo  {journal} {Phys. Rev. D}\ }\textbf {\bibinfo {volume}
  {89}},\ \bibinfo {pages} {054027} (\bibinfo {year} {2014})}\BibitemShut
  {NoStop}%
\bibitem [{\citenamefont {Herbst}\ \emph {et~al.}(2014)\citenamefont {Herbst},
  \citenamefont {Mitter}, \citenamefont {Pawlowski}, \citenamefont {Schaefer},\
  and\ \citenamefont {Stiele}}]{Herbst:2013ufa}%
  \BibitemOpen
  \bibfield  {author} {\bibinfo {author} {\bibfnamefont {T.~K.}\ \bibnamefont
  {Herbst}}, \bibinfo {author} {\bibfnamefont {M.}~\bibnamefont {Mitter}},
  \bibinfo {author} {\bibfnamefont {J.~M.}\ \bibnamefont {Pawlowski}}, \bibinfo
  {author} {\bibfnamefont {B.-J.}\ \bibnamefont {Schaefer}}, \ and\ \bibinfo
  {author} {\bibfnamefont {R.}~\bibnamefont {Stiele}},\ }\href {\doibase
  10.1016/j.physletb.2014.02.045} {\bibfield  {journal} {\bibinfo  {journal}
  {Phys. Lett. B}\ }\textbf {\bibinfo {volume} {731}},\ \bibinfo {pages} {248}
  (\bibinfo {year} {2014})}\BibitemShut {NoStop}%
\bibitem [{\citenamefont {Amelino-Camelia}\ and\ \citenamefont
  {Pi}(1993)}]{AmelinoCamelia:1992nc}%
  \BibitemOpen
  \bibfield  {author} {\bibinfo {author} {\bibfnamefont {G.}~\bibnamefont
  {Amelino-Camelia}}\ and\ \bibinfo {author} {\bibfnamefont {S.-Y.}\
  \bibnamefont {Pi}},\ }\href {\doibase 10.1103/PhysRevD.47.2356} {\bibfield
  {journal} {\bibinfo  {journal} {Phys. Rev. D}\ }\textbf {\bibinfo {volume}
  {47}},\ \bibinfo {pages} {2356} (\bibinfo {year} {1993})}\BibitemShut
  {NoStop}%
\bibitem [{\citenamefont {Amelino-Camelia}(1997)}]{AmelinoCamelia:1997dd}%
  \BibitemOpen
  \bibfield  {author} {\bibinfo {author} {\bibfnamefont {G.}~\bibnamefont
  {Amelino-Camelia}},\ }\href {\doibase 10.1016/S0370-2693(97)00709-0}
  {\bibfield  {journal} {\bibinfo  {journal} {Phys. Lett. B}\ }\textbf
  {\bibinfo {volume} {407}},\ \bibinfo {pages} {268} (\bibinfo {year}
  {1997})}\BibitemShut {NoStop}%
\bibitem [{\citenamefont {Petropoulos}(1999)}]{Petropoulos:1998gt}%
  \BibitemOpen
  \bibfield  {author} {\bibinfo {author} {\bibfnamefont {N.}~\bibnamefont
  {Petropoulos}},\ }\href {\doibase 10.1088/0954-3899/25/11/305} {\bibfield
  {journal} {\bibinfo  {journal} {J. Phys. G}\ }\textbf {\bibinfo {volume}
  {25}},\ \bibinfo {pages} {2225} (\bibinfo {year} {1999})}\BibitemShut
  {NoStop}%
\bibitem [{\citenamefont {Lenaghan}\ and\ \citenamefont
  {Rischke}(2000)}]{Lenaghan:1999si}%
  \BibitemOpen
  \bibfield  {author} {\bibinfo {author} {\bibfnamefont {J.~T.}\ \bibnamefont
  {Lenaghan}}\ and\ \bibinfo {author} {\bibfnamefont {D.~H.}\ \bibnamefont
  {Rischke}},\ }\href {\doibase 10.1088/0954-3899/26/4/309} {\bibfield
  {journal} {\bibinfo  {journal} {J. Phys. G}\ }\textbf {\bibinfo {volume}
  {26}},\ \bibinfo {pages} {431} (\bibinfo {year} {2000})}\BibitemShut
  {NoStop}%
\bibitem [{\citenamefont {Lenaghan}\ \emph {et~al.}(2000)\citenamefont
  {Lenaghan}, \citenamefont {Rischke},\ and\ \citenamefont
  {Schaffner-Bielich}}]{Lenaghan:2000ey}%
  \BibitemOpen
  \bibfield  {author} {\bibinfo {author} {\bibfnamefont {J.~T.}\ \bibnamefont
  {Lenaghan}}, \bibinfo {author} {\bibfnamefont {D.~H.}\ \bibnamefont
  {Rischke}}, \ and\ \bibinfo {author} {\bibfnamefont {J.}~\bibnamefont
  {Schaffner-Bielich}},\ }\href {\doibase 10.1103/PhysRevD.62.085008}
  {\bibfield  {journal} {\bibinfo  {journal} {Phys. Rev. D}\ }\textbf {\bibinfo
  {volume} {62}},\ \bibinfo {pages} {085008} (\bibinfo {year}
  {2000})}\BibitemShut {NoStop}%
\bibitem [{\citenamefont {Jakovac}\ \emph {et~al.}(2004)\citenamefont
  {Jakovac}, \citenamefont {Patkos}, \citenamefont {Szep},\ and\ \citenamefont
  {Szepfalusy}}]{Jakovac:2003ar}%
  \BibitemOpen
  \bibfield  {author} {\bibinfo {author} {\bibfnamefont {A.}~\bibnamefont
  {Jakovac}}, \bibinfo {author} {\bibfnamefont {A.}~\bibnamefont {Patkos}},
  \bibinfo {author} {\bibfnamefont {Z.}~\bibnamefont {Szep}}, \ and\ \bibinfo
  {author} {\bibfnamefont {P.}~\bibnamefont {Szepfalusy}},\ }\href {\doibase
  10.1016/j.physletb.2004.01.008} {\bibfield  {journal} {\bibinfo  {journal}
  {Phys. Lett. B}\ }\textbf {\bibinfo {volume} {582}},\ \bibinfo {pages} {179}
  (\bibinfo {year} {2004})}\BibitemShut {NoStop}%
\bibitem [{\citenamefont {Andersen}\ \emph {et~al.}(2004)\citenamefont
  {Andersen}, \citenamefont {Boer},\ and\ \citenamefont
  {Warringa}}]{Andersen:2004ae}%
  \BibitemOpen
  \bibfield  {author} {\bibinfo {author} {\bibfnamefont {J.~O.}\ \bibnamefont
  {Andersen}}, \bibinfo {author} {\bibfnamefont {D.}~\bibnamefont {Boer}}, \
  and\ \bibinfo {author} {\bibfnamefont {H.~J.}\ \bibnamefont {Warringa}},\
  }\href {\doibase 10.1103/PhysRevD.70.116007} {\bibfield  {journal} {\bibinfo
  {journal} {Phys. Rev. D}\ }\textbf {\bibinfo {volume} {70}},\ \bibinfo
  {pages} {116007} (\bibinfo {year} {2004})}\BibitemShut {NoStop}%
\bibitem [{\citenamefont {Andersen}\ and\ \citenamefont
  {Brauner}(2008)}]{Andersen:2008qk}%
  \BibitemOpen
  \bibfield  {author} {\bibinfo {author} {\bibfnamefont {J.~O.}\ \bibnamefont
  {Andersen}}\ and\ \bibinfo {author} {\bibfnamefont {T.}~\bibnamefont
  {Brauner}},\ }\href {\doibase 10.1103/PhysRevD.78.014030} {\bibfield
  {journal} {\bibinfo  {journal} {Phys. Rev. D}\ }\textbf {\bibinfo {volume}
  {78}},\ \bibinfo {pages} {014030} (\bibinfo {year} {2008})}\BibitemShut
  {NoStop}%
\bibitem [{\citenamefont {Nambu}\ and\ \citenamefont
  {Jona-Lasinio}(1961{\natexlab{a}})}]{Nambu:1961tp}%
  \BibitemOpen
  \bibfield  {author} {\bibinfo {author} {\bibfnamefont {Y.}~\bibnamefont
  {Nambu}}\ and\ \bibinfo {author} {\bibfnamefont {G.}~\bibnamefont
  {Jona-Lasinio}},\ }\href {\doibase 10.1103/PhysRev.122.345} {\bibfield
  {journal} {\bibinfo  {journal} {Phys. Rev.}\ }\textbf {\bibinfo {volume}
  {122}},\ \bibinfo {pages} {345} (\bibinfo {year}
  {1961}{\natexlab{a}})}\BibitemShut {NoStop}%
\bibitem [{\citenamefont {Nambu}\ and\ \citenamefont
  {Jona-Lasinio}(1961{\natexlab{b}})}]{Nambu:1961fr}%
  \BibitemOpen
  \bibfield  {author} {\bibinfo {author} {\bibfnamefont {Y.}~\bibnamefont
  {Nambu}}\ and\ \bibinfo {author} {\bibfnamefont {G.}~\bibnamefont
  {Jona-Lasinio}},\ }\href {\doibase 10.1103/PhysRev.124.246} {\bibfield
  {journal} {\bibinfo  {journal} {Phys. Rev.}\ }\textbf {\bibinfo {volume}
  {124}},\ \bibinfo {pages} {246} (\bibinfo {year}
  {1961}{\natexlab{b}})}\BibitemShut {NoStop}%
\bibitem [{\citenamefont {Fukushima}(2004)}]{Fukushima:2003fw}%
  \BibitemOpen
  \bibfield  {author} {\bibinfo {author} {\bibfnamefont {K.}~\bibnamefont
  {Fukushima}},\ }\href {\doibase 10.1016/j.physletb.2004.04.027} {\bibfield
  {journal} {\bibinfo  {journal} {Phys. Lett. B}\ }\textbf {\bibinfo {volume}
  {591}},\ \bibinfo {pages} {277} (\bibinfo {year} {2004})}\BibitemShut
  {NoStop}%
\bibitem [{\citenamefont {Ratti}\ \emph {et~al.}(2006)\citenamefont {Ratti},
  \citenamefont {Thaler},\ and\ \citenamefont {Weise}}]{Ratti:2005jh}%
  \BibitemOpen
  \bibfield  {author} {\bibinfo {author} {\bibfnamefont {C.}~\bibnamefont
  {Ratti}}, \bibinfo {author} {\bibfnamefont {M.~A.}\ \bibnamefont {Thaler}}, \
  and\ \bibinfo {author} {\bibfnamefont {W.}~\bibnamefont {Weise}},\ }\href
  {\doibase 10.1103/PhysRevD.73.014019} {\bibfield  {journal} {\bibinfo
  {journal} {Phys. Rev. D}\ }\textbf {\bibinfo {volume} {73}},\ \bibinfo
  {pages} {014019} (\bibinfo {year} {2006})}\BibitemShut {NoStop}%
\bibitem [{\citenamefont {Sasaki}\ \emph {et~al.}(2007)\citenamefont {Sasaki},
  \citenamefont {Friman},\ and\ \citenamefont {Redlich}}]{Sasaki:2006ww}%
  \BibitemOpen
  \bibfield  {author} {\bibinfo {author} {\bibfnamefont {C.}~\bibnamefont
  {Sasaki}}, \bibinfo {author} {\bibfnamefont {B.}~\bibnamefont {Friman}}, \
  and\ \bibinfo {author} {\bibfnamefont {K.}~\bibnamefont {Redlich}},\ }\href
  {\doibase 10.1103/PhysRevD.75.074013} {\bibfield  {journal} {\bibinfo
  {journal} {Phys. Rev. D}\ }\textbf {\bibinfo {volume} {75}},\ \bibinfo
  {pages} {074013} (\bibinfo {year} {2007})}\BibitemShut {NoStop}%
\bibitem [{\citenamefont {Roessner}\ \emph {et~al.}(2007)\citenamefont
  {Roessner}, \citenamefont {Ratti},\ and\ \citenamefont
  {Weise}}]{Roessner:2006xn}%
  \BibitemOpen
  \bibfield  {author} {\bibinfo {author} {\bibfnamefont {S.}~\bibnamefont
  {Roessner}}, \bibinfo {author} {\bibfnamefont {C.}~\bibnamefont {Ratti}}, \
  and\ \bibinfo {author} {\bibfnamefont {W.}~\bibnamefont {Weise}},\ }\href
  {\doibase 10.1103/PhysRevD.75.034007} {\bibfield  {journal} {\bibinfo
  {journal} {Phys. Rev. D}\ }\textbf {\bibinfo {volume} {75}},\ \bibinfo
  {pages} {034007} (\bibinfo {year} {2007})}\BibitemShut {NoStop}%
\bibitem [{\citenamefont {Fukushima}(2008)}]{Fukushima:2008wg}%
  \BibitemOpen
  \bibfield  {author} {\bibinfo {author} {\bibfnamefont {K.}~\bibnamefont
  {Fukushima}},\ }\href {\doibase 10.1103/PhysRevD.77.114028,
  10.1103/PhysRevD.78.039902} {\bibfield  {journal} {\bibinfo  {journal} {Phys.
  Rev. D}\ }\textbf {\bibinfo {volume} {77}},\ \bibinfo {pages} {114028}
  (\bibinfo {year} {2008})},\ \bibinfo {note} {[Erratum: Phys.
  Rev.D78,039902(2008)]}\BibitemShut {NoStop}%
\bibitem [{\citenamefont {Wetterich}(1993)}]{Wetterich:1992yh}%
  \BibitemOpen
  \bibfield  {author} {\bibinfo {author} {\bibfnamefont {C.}~\bibnamefont
  {Wetterich}},\ }\href {\doibase 10.1016/0370-2693(93)90726-X} {\bibfield
  {journal} {\bibinfo  {journal} {Phys. Lett. B}\ }\textbf {\bibinfo {volume}
  {301}},\ \bibinfo {pages} {90} (\bibinfo {year} {1993})}\BibitemShut
  {NoStop}%
\bibitem [{\citenamefont {Morris}(1994)}]{Morris:1993qb}%
  \BibitemOpen
  \bibfield  {author} {\bibinfo {author} {\bibfnamefont {T.~R.}\ \bibnamefont
  {Morris}},\ }\href {\doibase 10.1142/S0217751X94000972} {\bibfield  {journal}
  {\bibinfo  {journal} {Int. J. Mod. Phys. A}\ }\textbf {\bibinfo {volume}
  {09}},\ \bibinfo {pages} {2411} (\bibinfo {year} {1994})}\BibitemShut
  {NoStop}%
\bibitem [{\citenamefont {Berges}\ \emph {et~al.}(2002)\citenamefont {Berges},
  \citenamefont {Tetradis},\ and\ \citenamefont {Wetterich}}]{Berges:2000ew}%
  \BibitemOpen
  \bibfield  {author} {\bibinfo {author} {\bibfnamefont {J.}~\bibnamefont
  {Berges}}, \bibinfo {author} {\bibfnamefont {N.}~\bibnamefont {Tetradis}}, \
  and\ \bibinfo {author} {\bibfnamefont {C.}~\bibnamefont {Wetterich}},\ }\href
  {\doibase 10.1016/S0370-1573(01)00098-9} {\bibfield  {journal} {\bibinfo
  {journal} {Phys. Rept.}\ }\textbf {\bibinfo {volume} {363}},\ \bibinfo
  {pages} {223} (\bibinfo {year} {2002})}\BibitemShut {NoStop}%
\bibitem [{\citenamefont {Polonyi}(2003)}]{Polonyi:2001se}%
  \BibitemOpen
  \bibfield  {author} {\bibinfo {author} {\bibfnamefont {J.}~\bibnamefont
  {Polonyi}},\ }\href {\doibase 10.2478/BF02475552} {\bibfield  {journal}
  {\bibinfo  {journal} {Central Eur. J. Phys.}\ }\textbf {\bibinfo {volume}
  {1}},\ \bibinfo {pages} {1} (\bibinfo {year} {2003})}\BibitemShut {NoStop}%
\bibitem [{\citenamefont {Braun}\ \emph {et~al.}(2011)\citenamefont {Braun},
  \citenamefont {Klein},\ and\ \citenamefont {Piasecki}}]{Braun:2010vd}%
  \BibitemOpen
  \bibfield  {author} {\bibinfo {author} {\bibfnamefont {J.}~\bibnamefont
  {Braun}}, \bibinfo {author} {\bibfnamefont {B.}~\bibnamefont {Klein}}, \ and\
  \bibinfo {author} {\bibfnamefont {P.}~\bibnamefont {Piasecki}},\ }\href
  {\doibase 10.1140/epjc/s10052-011-1576-7} {\bibfield  {journal} {\bibinfo
  {journal} {Eur. Phys. J. C}\ }\textbf {\bibinfo {volume} {71}},\ \bibinfo
  {pages} {1576} (\bibinfo {year} {2011})}\BibitemShut {NoStop}%
\bibitem [{\citenamefont {Landau}\ and\ \citenamefont
  {Lifshitz}(1980)}]{landau2013statistical}%
  \BibitemOpen
  \bibfield  {author} {\bibinfo {author} {\bibfnamefont {L.}~\bibnamefont
  {Landau}}\ and\ \bibinfo {author} {\bibfnamefont {E.}~\bibnamefont
  {Lifshitz}},\ }\href@noop {} {\emph {\bibinfo {title} {Statistical Physics,
  Part~1}}},\ \bibinfo {number} {Course of Theoretical Physics, Vol. 5, 3rd
  ed.}\ (\bibinfo  {publisher} {Butterworth-Heinemann},\ \bibinfo {year}
  {1980})\BibitemShut {NoStop}%
\bibitem [{\citenamefont {Koch}(1997)}]{Koch:1997ei}%
  \BibitemOpen
  \bibfield  {author} {\bibinfo {author} {\bibfnamefont {V.}~\bibnamefont
  {Koch}},\ }\href {\doibase 10.1142/S0218301397000147} {\bibfield  {journal}
  {\bibinfo  {journal} {Int. J. Mod. Phys. E}\ }\textbf {\bibinfo {volume}
  {06}},\ \bibinfo {pages} {203} (\bibinfo {year} {1997})}\BibitemShut
  {NoStop}%
\bibitem [{\citenamefont {Dolan}\ and\ \citenamefont
  {Jackiw}(1974)}]{Dolan:1973qd}%
  \BibitemOpen
  \bibfield  {author} {\bibinfo {author} {\bibfnamefont {L.}~\bibnamefont
  {Dolan}}\ and\ \bibinfo {author} {\bibfnamefont {R.}~\bibnamefont {Jackiw}},\
  }\href {\doibase 10.1103/PhysRevD.9.3320} {\bibfield  {journal} {\bibinfo
  {journal} {Phys. Rev. D}\ }\textbf {\bibinfo {volume} {9}},\ \bibinfo {pages}
  {3320} (\bibinfo {year} {1974})}\BibitemShut {NoStop}%
\bibitem [{\citenamefont {Klajn}(2014)}]{Klajn:2013gxa}%
  \BibitemOpen
  \bibfield  {author} {\bibinfo {author} {\bibfnamefont {B.}~\bibnamefont
  {Klajn}},\ }\href {\doibase 10.1103/PhysRevD.89.036001} {\bibfield  {journal}
  {\bibinfo  {journal} {Phys. Rev. D}\ }\textbf {\bibinfo {volume} {89}},\
  \bibinfo {pages} {036001} (\bibinfo {year} {2014})}\BibitemShut {NoStop}%
\bibitem [{\citenamefont {Fukushima}(2003{\natexlab{a}})}]{Fukushima:2002ew}%
  \BibitemOpen
  \bibfield  {author} {\bibinfo {author} {\bibfnamefont {K.}~\bibnamefont
  {Fukushima}},\ }\href {\doibase 10.1016/S0370-2693(02)03184-2} {\bibfield
  {journal} {\bibinfo  {journal} {Phys. Lett. B}\ }\textbf {\bibinfo {volume}
  {553}},\ \bibinfo {pages} {38} (\bibinfo {year}
  {2003}{\natexlab{a}})}\BibitemShut {NoStop}%
\bibitem [{\citenamefont {Fukushima}(2003{\natexlab{b}})}]{Fukushima:2003fm}%
  \BibitemOpen
  \bibfield  {author} {\bibinfo {author} {\bibfnamefont {K.}~\bibnamefont
  {Fukushima}},\ }\href {\doibase 10.1103/PhysRevD.68.045004} {\bibfield
  {journal} {\bibinfo  {journal} {Phys. Rev. D}\ }\textbf {\bibinfo {volume}
  {68}},\ \bibinfo {pages} {045004} (\bibinfo {year}
  {2003}{\natexlab{b}})}\BibitemShut {NoStop}%
\bibitem [{\citenamefont {Lo}\ \emph {et~al.}(2013)\citenamefont {Lo},
  \citenamefont {Friman}, \citenamefont {Kaczmarek}, \citenamefont {Redlich},\
  and\ \citenamefont {Sasaki}}]{Lo:2013hla}%
  \BibitemOpen
  \bibfield  {author} {\bibinfo {author} {\bibfnamefont {P.~M.}\ \bibnamefont
  {Lo}}, \bibinfo {author} {\bibfnamefont {B.}~\bibnamefont {Friman}}, \bibinfo
  {author} {\bibfnamefont {O.}~\bibnamefont {Kaczmarek}}, \bibinfo {author}
  {\bibfnamefont {K.}~\bibnamefont {Redlich}}, \ and\ \bibinfo {author}
  {\bibfnamefont {C.}~\bibnamefont {Sasaki}},\ }\href {\doibase
  10.1103/PhysRevD.88.074502} {\bibfield  {journal} {\bibinfo  {journal} {Phys.
  Rev. D}\ }\textbf {\bibinfo {volume} {88}},\ \bibinfo {pages} {074502}
  (\bibinfo {year} {2013})}\BibitemShut {NoStop}%
\bibitem [{Note1()}]{Note1}%
  \BibitemOpen
  \bibinfo {note} {The stationary point is a saddle point in the variables
  $\Phi $ and $\protect \mathaccentV {bar}016{\Phi }$. Nevertheless, for the
  effective Polyakov loop potential employed in this paper, the system is
  thermodynamically stable, as shown in ref.~\cite
  {Sasaki:2006ww}.}\BibitemShut {Stop}%
\bibitem [{\citenamefont {Baym}(1962)}]{Baym:1962sx}%
  \BibitemOpen
  \bibfield  {author} {\bibinfo {author} {\bibfnamefont {G.}~\bibnamefont
  {Baym}},\ }\href {\doibase 10.1103/PhysRev.127.1391} {\bibfield  {journal}
  {\bibinfo  {journal} {Phys. Rev.}\ }\textbf {\bibinfo {volume} {127}},\
  \bibinfo {pages} {1391} (\bibinfo {year} {1962})}\BibitemShut {NoStop}%
\bibitem [{\citenamefont {Cornwall}\ \emph {et~al.}(1974)\citenamefont
  {Cornwall}, \citenamefont {Jackiw},\ and\ \citenamefont
  {Tomboulis}}]{Cornwall:1974vz}%
  \BibitemOpen
  \bibfield  {author} {\bibinfo {author} {\bibfnamefont {J.~M.}\ \bibnamefont
  {Cornwall}}, \bibinfo {author} {\bibfnamefont {R.}~\bibnamefont {Jackiw}}, \
  and\ \bibinfo {author} {\bibfnamefont {E.}~\bibnamefont {Tomboulis}},\ }\href
  {\doibase 10.1103/PhysRevD.10.2428} {\bibfield  {journal} {\bibinfo
  {journal} {Phys. Rev. D}\ }\textbf {\bibinfo {volume} {10}},\ \bibinfo
  {pages} {2428} (\bibinfo {year} {1974})}\BibitemShut {NoStop}%
\bibitem [{\citenamefont {van Hees}\ and\ \citenamefont
  {Knoll}(2001)}]{vanHees:2001ik}%
  \BibitemOpen
  \bibfield  {author} {\bibinfo {author} {\bibfnamefont {H.}~\bibnamefont {van
  Hees}}\ and\ \bibinfo {author} {\bibfnamefont {J.}~\bibnamefont {Knoll}},\
  }\href {\doibase 10.1103/PhysRevD.65.025010} {\bibfield  {journal} {\bibinfo
  {journal} {Phys. Rev. D}\ }\textbf {\bibinfo {volume} {65}},\ \bibinfo
  {pages} {025010} (\bibinfo {year} {2001})}\BibitemShut {NoStop}%
\bibitem [{\citenamefont {Van~Hees}\ and\ \citenamefont
  {Knoll}(2002)}]{VanHees:2001pf}%
  \BibitemOpen
  \bibfield  {author} {\bibinfo {author} {\bibfnamefont {H.}~\bibnamefont
  {Van~Hees}}\ and\ \bibinfo {author} {\bibfnamefont {J.}~\bibnamefont
  {Knoll}},\ }\href {\doibase 10.1103/PhysRevD.65.105005} {\bibfield  {journal}
  {\bibinfo  {journal} {Phys. Rev. D}\ }\textbf {\bibinfo {volume} {65}},\
  \bibinfo {pages} {105005} (\bibinfo {year} {2002})}\BibitemShut {NoStop}%
\bibitem [{\citenamefont {van Hees}\ and\ \citenamefont
  {Knoll}(2002)}]{vanHees:2002bv}%
  \BibitemOpen
  \bibfield  {author} {\bibinfo {author} {\bibfnamefont {H.}~\bibnamefont {van
  Hees}}\ and\ \bibinfo {author} {\bibfnamefont {J.}~\bibnamefont {Knoll}},\
  }\href {\doibase 10.1103/PhysRevD.66.025028} {\bibfield  {journal} {\bibinfo
  {journal} {Phys. Rev. D}\ }\textbf {\bibinfo {volume} {66}},\ \bibinfo
  {pages} {025028} (\bibinfo {year} {2002})}\BibitemShut {NoStop}%
\bibitem [{\citenamefont {Baxter}(1982)}]{Baxter:1982zz}%
  \BibitemOpen
  \bibfield  {author} {\bibinfo {author} {\bibfnamefont {R.~J.}\ \bibnamefont
  {Baxter}},\ }\href@noop {} {\emph {\bibinfo {title} {{Exactly Solved Models
  in Statistical Mechanics}}}}\ (\bibinfo  {publisher} {Academic Press Inc.},\
  \bibinfo {year} {London, 1982})\BibitemShut {NoStop}%
\bibitem [{\citenamefont {Engels}\ and\ \citenamefont
  {Karsch}(2014)}]{Engels:2014bra}%
  \BibitemOpen
  \bibfield  {author} {\bibinfo {author} {\bibfnamefont {J.}~\bibnamefont
  {Engels}}\ and\ \bibinfo {author} {\bibfnamefont {F.}~\bibnamefont
  {Karsch}},\ }\href {\doibase 10.1103/PhysRevD.90.014501} {\bibfield
  {journal} {\bibinfo  {journal} {Phys. Rev. D}\ }\textbf {\bibinfo {volume}
  {90}},\ \bibinfo {pages} {014501} (\bibinfo {year} {2014})}\BibitemShut
  {NoStop}%
\bibitem [{\citenamefont {Bohr}\ \emph {et~al.}(2001)\citenamefont {Bohr},
  \citenamefont {Schaefer},\ and\ \citenamefont {Wambach}}]{Bohr:2000gp}%
  \BibitemOpen
  \bibfield  {author} {\bibinfo {author} {\bibfnamefont {O.}~\bibnamefont
  {Bohr}}, \bibinfo {author} {\bibfnamefont {B.~J.}\ \bibnamefont {Schaefer}},
  \ and\ \bibinfo {author} {\bibfnamefont {J.}~\bibnamefont {Wambach}},\ }\href
  {\doibase 10.1142/S0217751X0100502X} {\bibfield  {journal} {\bibinfo
  {journal} {Int. J. Mod. Phys. A}\ }\textbf {\bibinfo {volume} {16}},\
  \bibinfo {pages} {3823} (\bibinfo {year} {2001})}\BibitemShut {NoStop}%
\bibitem [{\citenamefont {Borsanyi}\ \emph {et~al.}(2010)\citenamefont
  {Borsanyi} \emph {et~al.}}]{Borsanyi:2010bp}%
  \BibitemOpen
  \bibfield  {author} {\bibinfo {author} {\bibfnamefont {S.}~\bibnamefont
  {Borsanyi}} \emph {et~al.} (\bibinfo {collaboration} {Wuppertal-Budapest}),\
  }\href {\doibase 10.1007/JHEP09(2010)073} {\bibfield  {journal} {\bibinfo
  {journal} {JHEP}\ }\textbf {\bibinfo {volume} {1009}},\ \bibinfo {pages}
  {073} (\bibinfo {year} {2010})}\BibitemShut {NoStop}%
\end{thebibliography}%

\end{document}